\documentstyle[12pt,epsf]{article}

\newcommand{\bmat}{\left(\begin{array}}
\newcommand{\emat}{\end{array}\right)}
\def\NPB#1#2#3{Nucl. Phys. B{#1} (19#2) #3}
\def\PLB#1#2#3{Phys. Lett. B{#1} (19#2) #3}

\def\PRD#1#2#3{Phys. Rev. D{#1} (19#2) #3}

\def\MODA#1#2#3{Mod. Phys. Lett.  {#1} (19#2) #3}

\def\yzero{\smash{\hbox{$y\kern-4pt\raise1pt\hbox{${}^\circ$}$}}}

\def\-{\hphantom{-}}
\def\ov{\overline}
\def\s2{\frac{1}{\sqrt2}}

\def\beq{\begin{equation}}
\def\eeq{\end{equation}}
\def\beqa{\begin{eqnarray}}
\def\eeqa{\end{eqnarray}}
\def\tr{{\rm tr \,}}
\def\Tr{{\rm Tr \,}}
\def\diag{{\rm diag \,}}

\def\IF{\relax{\rm I\kern-.18em F}}
\def\II{\relax{\rm I\kern-.18em I}}
\def\IP{\relax{\rm I\kern-.18em P}}

\def\NN{{\cal N}}

\def\rr{{\cal R}}

\def\Dsl{\,\raise.15ex\hbox{/}\mkern-13.5mu D} 

\def\IC{\bf C}
\def\IZ{\bf Z}

\def\z2z2{$\IC^3/(\IZ_2\times\IZ_2)$}
\def\twothree{$\IC^3/(\IZ_2\times\IZ_3)$ }

\def\nsp{NS$^{(\;)}$\kern-.55em $'\;\,$}

\newcommand{\drawsquare}[2]{\hbox{%
\rule{#2pt}{#1pt}\hskip-#2pt
\rule{#1pt}{#2pt}\hskip-#1pt
\rule[#1pt]{#1pt}{#2pt}}\rule[#1pt]{#2pt}{#2pt}\hskip-#2pt
\rule{#2pt}{#1pt}}

\newcommand{\fund}{\raisebox{-.5pt}{\drawsquare{6.5}{0.4}}}
\newcommand{\Ysymm}{\raisebox{-.5pt}{\drawsquare{6.5}{0.4}}\hskip-0.4pt%
        \raisebox{-.5pt}{\drawsquare{6.5}{0.4}}}
\newcommand{\Yasymm}{\raisebox{-3.5pt}{\drawsquare{6.5}{0.4}}\hskip-6.9pt%
        \raisebox{3pt}{\drawsquare{6.5}{0.4}}}
\newcommand{\antifund}{\overline{\fund}}
\newcommand{\bYasymm}{\overline{\Yasymm}}

\newcommand{\smsymm}{\raisebox{-.5pt}{\drawsquare{3.5}{0.4}}\hskip-0.4pt%
        \raisebox{-.5pt}{\drawsquare{3.5}{0.4}}}
\newcommand{\smasymm}{\raisebox{-3.5pt}{\drawsquare{3.5}{0.4}}\hskip-3.9pt%
        \raisebox{0pt}{\drawsquare{3.5}{0.4}}}

\topmargin
-1.5cm
\textwidth
15.5cm
\textheight
24cm
\oddsidemargin
0.7cm
\evensidemargin
1.2cm

\begin{document}

\makeatletter
\@addtoreset{equation}{section}
\makeatother
\renewcommand{\theequation}{\thesection.\arabic{equation}}
\pagestyle{empty}
\rightline{FTUAM-99/22}
\rightline{IASSNS-HEP-99/62}
\rightline{IFT-UAM/CSIC-99-26}

\rightline{\tt hep-th/9907086}
\vspace{0.5cm}
\begin{center}
\LARGE{Orientifolding the conifold \\[10mm]}
\large{
J.~Park$^\dagger$\footnote{\texttt{jaemo@ias.edu}},
R. Rabad\'an$^\S$\footnote{\texttt{rabadan@delta.ft.uam.es}} and
A.~M.~Uranga$^\dagger$\footnote{\texttt{uranga@ias.edu}}\\[2mm]}
\small{
$^\dagger$ {\em School of Natural Sciences, Institute for Advanced Study, 
\\ Olden Lane, Princeton NJ 08540, USA}\\[4mm]
$^\S$ {\em Departamento de F\'{\i}sica Te\'orica C-XI
and Instituto de F\'{\i}sica Te\'orica  C-XVI,\\[-0.3em]
Universidad Aut\'onoma de Madrid,
Cantoblanco, 28049 Madrid, Spain}\\[4mm]
}
\small{\bf Abstract} \\[7mm]
\end{center}

\begin{center}
\begin{minipage}[h]{14.0cm}

{\small In this paper we study the $\NN=1$ supersymmetric field theories 
realized on the world-volume of type IIB D3-branes sitting at orientifolds 
of non-orbifold singularities (conifold and generalizations). Several 
chiral models belong to this family of theories. These field theories 
have a T-dual realization in terms of type IIA configurations of 
relatively rotated NS fivebranes, D4-branes and orientifold six-planes, 
with a compact $x^6$ direction, along which the D4-branes have finite 
extent. We compute the spectrum on the D3-branes directly in the type IIB 
picture and match the resulting field theories with those obtained in the 
type IIA setup, thus providing a non-trivial check of this T-duality. 
Since the usual techniques to compute the spectrum of the model and check 
the cancellation of tadpoles, cannot be applied to the case orientifolds 
of non-orbifold singularities, we use a different approach, and construct 
the models by partially blowing-up orientifolds of $\IC^3/(\IZ_2\times \IZ_2)$ 
and $\IC^3/(\IZ_2\times \IZ_3)$ orbifolds.
}

\end{minipage}
\end{center}

\newpage
\setcounter{page}{1}
\pagestyle{plain}
\renewcommand{\thefootnote}{\arabic{footnote}}
\setcounter{footnote}{0}

\section{Introduction}

The recent developments in the study of supersymmetric field theories by
embedding them into string theory has profited from the interplay of
different approaches. The most relevant ones for our purposes in this
paper are the realization in terms of D-branes in the presence of
NS-branes (and possibly other higher-dimensional D-branes)
\cite{hw,egk,review,hzbb}, and in terms of D-branes probes at spacetime
singularities \cite{dm,dgm}. These two broad classes of constructions are
related by T-dualities \cite{kls,hanur,uraconi,keshav}, which transform the
NS-fivebranes into geometric singularities \cite{oovafa}. This type of
mapping allows to answer different questions in the different pictures.
Thus, the brane configurations of NS-branes and D-branes are very
intuitive and allow to easily classify large classes of models and their
corresponding field theories. On the other hand, the picture of branes at
singularities contains only perturbative objects and is better suited for
some explicit computations of perturbative effects in the corresponding
field theory, for instance anomaly cancellation conditions and one-loop
beta functions \cite{lr,iru}. Also, the large N limit of the field
theory can
be studied in term of a dual supergravity (or superstring) background by
means of the AdS/CFT correspondence \cite{revads}.

In this paper we are going to consider type IIA configurations of
NS-fivebranes (with world-volume along the directions 012345),
NS$'$-branes (along 012389), with D4-branes (along 01236) suspended
between them, and in the presence of D6$'$-branes and O6$'$-planes (both
along 0123457). These constructions realize a variety of four-dimensional 
$\NN=1$ supersymmetric gauge field theories on the D4-brane world-volume. 
Configurations of this type, but without orientifold planes, were first 
introduced in \cite{egk}. The introductions of orientifold planes was 
discussed in \cite{lll,egkt,bhkl} (see also \cite{park,lll2}).

We will be interested in configurations where the direction $x^6$, along
which the D4-branes are suspended between the NS fivebranes, is
compactified on a circle. Our basic aim is to understand the resulting
configurations after performing a T-duality along this direction. There
are basically two motivations for this. The first is that the IIB T-dual
picture provides an interesting insight into several non-trivial brane
dynamics effects that occur on the type IIA side, and which are related
to chiral symmetries and chiral matter in the gauge field theory. These
include the appearance of chiral symmetries and chiral flavours when a
half-D6$'$ brane ends on a NS-brane \cite{hanbro}, the appearance of
chiral matter due to the change of sign of the O6$'$-plane charge when it
crosses a NS-brane \cite{lll,egkt,bhkl}. The type IIB realization of these
effects has been studied in \cite{pru} (some had been previously observed
in \cite{kg}) in the simpler case where NS$'$-branes are absent
\footnote{Notice that our models have $\NN=1$ supersymmetry {\em before}
the orientifold projection, whereas \cite{pru} considered $\NN=1$
orientifolds of $\NN=2$ models.}. This paper can be regarded as an
extension of these result to more general models.

The second and maybe more interesting motivation is that the T-dual
configurations involve orientifolds of non-orbifold singularities. This
can be seen as follows. Type IIA configurations of $k$ NS-branes, $k'$
NS$'$-branes and D4-branes, without orientifold planes, but with compact
direction $x^6$, transform under T-duality into a set of D3-branes
probing the non-orbifold singularity $xy=z^{k'} {w}^k$ \cite{uraconi} (see
also \cite{keshav} for the particular case $k=k'=1$). Therefore, it is
reasonable to expect the IIA models with O6$'$-planes to correspond to
suitable orientifolds of these non-orbifold spaces. In this paper we would
like to construct such orientifolds directly on the IIB side and compare
the resulting field theories on the D3-brane probes with those obtained in
the IIA setup. Unfortunately, this problem is rather difficult, since
the usual techniques to construct type IIB orientifolds, compute the 
field theory spectrum and check the cancellation of twisted RR tadpoles 
(and ensure the consistency of the string theory configurations) 
\cite{sag1,dlp,horava,bs,gp,gjdp} do not apply, since the world-sheet
description of these models is not a solvable conformal field theory. On 
the
other hand, this means that the type IIA configurations can provide,
through the T-duality, interesting information about the possible
consistent orientifolds of non-orbifold spaces. In this paper, however, we
will construct the IIB models directly and use the results as new support
for the T-duality proposal.

The procedure we use is based in the observation in \cite{mp} that one can
construct non-orbifold singularities as partial resolutions of orbifold
singularities. The effect of the blow-ups appears in the D3-brane field
theory as specific Higgs breakings which can be identified in a precise
manner. The effective field theory along the Higgs branch gives the field
theory of the D3-branes at the non-orbifold singularity. We apply this
idea in the presence of an additional orientifold projection; that is, we
study partial resolutions of orientifolds of orbifold singularities to
construct orientifolds of non-orbifold singularities. Due to the
complexity of the method (concretely, the identification of the Higgsing
associated to a specific blow-up) we will mainly discuss the simplest
examples with at most three NS fivebranes. Some comments about more
general cases are mentioned at the end.

As mentioned above, the singularity realization
of these field theories provides a simple description of several exotic
phenomena in the IIA side. This point was already stressed in the simplest
context of \cite{pru}, so we will not insist on it here. Another
interesting observation suggested by our constructions is that the chiral
theories obtained in the IIA framework (and other related $\NN=1$ models)
are continuously connected to other chiral theories obtained by
orientifolding orbifold singularities. The relation is the
process of blowing-up of the singularity probed by the D3-branes, or
equivalently the Higgs breaking in the field theory. We find this picture
quite reassuring, and satisfactory, since it suggests a unified
description for all chiral gauge theories which can be embedded in
string theory.

\medskip

The paper is organized as follows. In Section~2 we review the field
theories arising from type IIA configurations with NS-branes, NS$'$-branes
and D4-branes, with the direction 6 compact, but without orientifold
planes. We also review the T-duality with sets of D3-branes at
non-orbifold singularities. As a preparation for analogous computations in
the orientifolded case, we review the construction in \cite{mp} of the
field theories on D3-branes at the conifold  ($xy=zw$) and suspended
pinch point ($xy=zw^2$) singularities, by blowing up the \z2z2 orbifold
singularity. The techniques of toric geometry required for these
computations have been confined to section 2.2.1, so that it can
be (hopefully) safely avoided by readers not interested in them.

In Section~3 we turn to the type IIA configurations including
O6$'$-planes. As explained above, we center on the simplest examples,
with one NS-brane and one NS$'$-brane (which T-dualize to orientifolds of
the conifold), or one NS-brane and two NS$'$-branes (which T-dualize to
orientifolds of the suspended pinch point).

The type IIB orientifolds of the suspended pinch point, $xy=zw^2$, are the
subject of Section~4. They can be obtained as partial resolutions of
suitable orientifolds of \z2z2. These are constructed in Section~4.1,
where we also check explicitly the cancellation of tadpoles. In
Section~4.2 we discuss the blow-ups which provide the orientifolds of
$xy=zw^2$, and the associated field theory Higgsings. The resulting field
theories match nicely those obtained from the type IIA constructions. We
also comment on some interesting results obtained upon continuing blowing-up.

In Section~5 we discuss orientifolds of the conifold. These cannot be
obtained by resolving orientifolds of \z2z2, since the orientifold
projection eliminates the required blow-up mode. However, they can be
constructed by resolving orientifolds of \twothree. We describe the
highlights of the computations involved, and describe the results. The
field theories obtained after the Higgsing again reproduce those obtained
from the IIA side.

Finally, in Section~6 we make several remarks concerning the
generalization of our results to arbitrary singularities $xy=z^kw^{k'}$
(T-duals of models with $k$ NS-branes and $k'$ NS$'$-branes, and end with
some final comments.

\medskip

\section{Rotated branes and non-orbifold singularities}

We start with a brief review of some simple IIA brane configurations, and
their IIB T-dual version as D3-branes at singularities, before the
introduction of orientifold projections.

\subsection{$\NN=1$ elliptic models}

There is a natural way to break the $\NN=2$ supersymmetry of the elliptic
models considered in \cite{wit4d} to $\NN=1$, namely to use NS-fivebranes
whose world-volumes are relatively rotated. In this paper we will consider
only fivebranes with worldvolume along 012345 (denoted NS-branes) and
along 012389 (denoted NS$'$-branes). Thus we restrict to fivebranes which
are either parallel or orthogonal, even when more general
angles are allowed. This type of configurations was first considered in
\cite{egk} (see \cite{review} for further references) in models with
non-compact $x^6$. We will be interested in models with $x^6$ compact, which
were discussed in \cite{uraconi,ehn}. We will refer to them as $\NN=1$ 
elliptic models.

It is straightforward to  read off the resulting $\NN=1$ field theories. For a
model with $k$ NS-branes and $l$ NS$'$-branes, we obtain a gauge group
$\prod_{i=1}^{k+l} SU(n_i)$ (times a decoupled $U(1)$, ignored in the
following), and bifundamental chiral multiplets $\sum_{i=1}^{k+l}\,
[\, (\fund_i,\antifund_{i+1}) + (\antifund_i,\fund_{i+1})\, ]$. We also get a
massless adjoint whenever two adjacent fivebranes are parallel, which
parametrizes a Coulomb branch in which D4-branes slide along the NS-branes.
When two adjacent fivebranes are orthogonal, there is quartic superpotential
for the bifundamental fields living at the ends of the interval (see
\cite{uraconi} for more details).

In \cite{uraconi} (and in \cite{keshav} in a particular case) it was argued
that a T-duality along $x^6$ maps this configuration to a system of D3-branes
at a non-orbifold singularity\footnote{For other references concerning
non-orbifold singularities (conifold and generalizations), see
\cite{kw,mp,gns,lopez,unge,ito,aklm,keshav2,tatar}} $xy=z^l w^k$. The result
follows from the T-duality between a NS fivebrane and a Taub-NUT space $xy=v$
\cite{oovafa}. In the general case, the singularity requires additional data
to specify the field theory completely. In particular, the different orderings
of NS and NS$'$ branes in $x^6$ corresponds to a specific choice of B-fields
in the collapsed two-cycles in the singularity. This subtlety will not arise
in the particular models we study.

\begin{figure}
\centering
\epsfxsize=4.5in
\hspace*{0in}\vspace*{.2in}
\epsffile{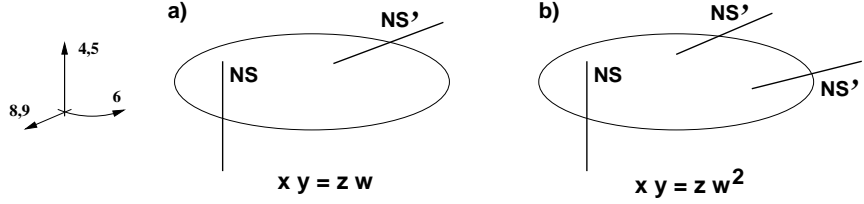}
\caption{\small Examples of $\NN=1$ elliptic models. Figure a) is T-dual
to a set of D3-branes at a conifold singularity $xy=zw$. Figure b) is
T-dual to D3-branes at the `suspended pinch point' singularity,
$xy=zw^2$.}
\label{fig:conspp}
\end{figure}
The main examples we will consider are depicted in figure \ref{fig:conspp}.
The first configuration, figure \ref{fig:conspp}a, contains one NS brane and
one NS$'$ brane. It realizes a field theory with the following gauge group
and $\NN=1$ matter content
{\small
\beqa
\begin{array}{cccc}
& SU(n_1) & SU(n_2) & \\
A_i & \fund & \antifund & i=1,2 \\
B_i & \antifund & \fund & i=1,2
\label{speconi}
\end{array}
\eeqa
}
The superpotential is given by
\beqa
W & = & \Tr(A_1B_1A_2B_2-A_1B_2A_2B_1).
\label{supconi}
\eeqa
After T-dualizing along $x^6$ this configuration maps to a set of D3
branes at a conifold singularity $xy=zw$. In fact, the above field theory
was proposed in \cite{kw} to arise on D3-branes at the conifold singularity,
on the basis of strong evidence from the AdS/CFT correspondence.

The second configuration, figure \ref{fig:conspp}b, contains one NS-brane
and two NS$'$-branes. The corresponding gauge group and matter content are
{\small
\beqa
\begin{array}{cccc}
& SU(n_1) & SU(n_2) & SU(n_3)\\
F & \fund & \antifund &  \\
{\tilde F} & \antifund & \fund &  \\
G & & \fund & \antifund   \\
{\tilde G} & & \antifund & \fund   \\
H & \antifund & & \fund \\
{\tilde H} & \fund & & \antifund \\
\Phi_1 & {\rm Adj.} & &
\end{array}
\label{specspp}
\eeqa
}
The superpotential is given by
\beqa
W & = & \tr({\tilde F}FG{\tilde G} - {\tilde G}G H{\tilde H} + {\tilde H}
H \Phi_1 -{\tilde F}\Phi_1 F).
\label{supspp}
\eeqa
After T-duality, this configuration transforms into a set of D3 branes at the
so-called `suspended pinch point' (SPP) singularity, $xy=zw^2$. This T-duality
is supported by the fact that independent arguments \cite{mp} (to be reviewed
in the following section) imply that the above field theory is indeed realized
on D3-branes at the SPP singularity.

It is rather difficult to study D3-branes at non-orbifold singularities,
the reason being that they are not described by free world-sheet conformal
field theories. From this point of view it is fortunate that for toric
singularities there exists a systematic (though involved) approach, based on
\cite{dgm} and developed in \cite{mp}, that allows to compute the spectrum and
interactions of the D3-brane field theory. Therefore, before entering the
study of the orientifold models, which are more complicated, it will be
convenient to review the description of the conifold and SPP models using
these techniques.

\subsection{Non-orbifold singularities from orbifold singularities}

The main observation is that the non-orbifold singularities of interest,
the conifold $xy=zw$ and the SPP $xy=zw^2$, can be constructed as partial
resolutions of the  \z2z2 orbifold singularity. The latter is obtained
upon modding out the $\IC^3$ parametrized by $(z_1,z_2,z_3)$ by the group
generated by
\beqa
\theta: & (z_1,z_2,z_3) \rightarrow (-z_1,z_2,-z_3) \nonumber\\
\omega: & (z_1,z_2,z_3) \rightarrow (z_1,-z_2,-z_3)
\label{actz2z2}
\eeqa
Defining invariant variables, $x=z_1^2$, $y=z_2^2$, $z=z_3^2$, $w=z_1 z_2
z_3$, the space can be describe as the hypersurface \vspace*{-.5cm}
\beqa
xyz=w^2
\eeqa
in $\IC^4$. This space can be blown-up once by introducing a $\IP_1$
parametrized by $w'=w/z$ (in the coordinate patch $z\neq 0$). The
remaining singularity has the form
\beqa
xy=zw'{\,^2}
\eeqa
which is a suspended pinch point singularity. There are two inequivalent ways
of performing a further blow-up. The first possibility is to introduce a
$\IP_1$ parametrized {\em e.g.} by $x'=x/w'$. The remaining singularity is
the conifold $x'y=zw'$. The second possibility is to introduce a $\IP_1$
parametrized {\em e.g.} by $y'=y/z$. The remaining singularity is $xy'=w'^2$,
the $\IC^2/\IZ_2\times \IC$ orbifold. For any of these two possibilities
further blow-ups yield a completely smooth space.

Our aim in this section is to construct the field theory of D3-branes at
these non-orbifold singularities by starting with the well-known system of
D3-branes at \z2z2 and following the effect of the above blow-ups in the
field theory.

\medskip

The field theory of a set of D3-branes at a \z2z2 singularity can be
easily constructed following \cite{dgm}. We mod out a system of D3
branes in flat space by the action (\ref{actz2z2}), embedded on the
D3-brane Chan-Paton factors through the matrices
\beqa
\gamma_{\theta,3} =\diag (1_{n_1},1_{n_2},-1_{n_3},-1_{n_4}) \nonumber \\
\gamma_{\omega,3} =\diag (1_{n_1},-1_{n_2},1_{n_3},-1_{n_4})
\eeqa
The resulting field theory is found by imposing the projections
{\small
\beqa
\begin{array}{cccc}
V=\gamma_{\theta,3} V \gamma_{\theta,3}^{-1} \; , &
X=-\gamma_{\theta,3} X \gamma_{\theta,3}^{-1} \; , &
Y=\gamma_{\theta,3} Y \gamma_{\theta,3}^{-1} \; ,&
Z=-\gamma_{\theta,3} Z \gamma_{\theta,3}^{-1}  \\
V=\gamma_{\omega,3} V \gamma_{\omega,3}^{-1} \; ,&
X=\gamma_{\omega,3} X \gamma_{\omega,3}^{-1} \; ,&
Y=-\gamma_{\omega,3} Y \gamma_{\omega,3}^{-1} \; ,&
Z=-\gamma_{\omega,3} Z \gamma_{\omega,3}^{-1}
\end{array}
\eeqa
}
The gauge group\footnote{We momentarily maintain the $U(1)$ factors in the
gauge group.} is $U(n_1)$$\times U(n_2)$$\times U(n_3)$$\times U(n_4)$, and
there are $\NN=1$ matter multiplets
{\small
\beqa
X_{13} : (\fund_1,\antifund_3) & \quad Y_{12} : (\fund_1,\antifund_2) &
\quad Z_{14} : (\fund_1,\antifund_4) \nonumber\\
X_{31} : (\fund_3,\antifund_1) & \quad Y_{21} : (\fund_2,\antifund_1) &
\quad Z_{41} : (\fund_4,\antifund_1) \nonumber\\
X_{24} : (\fund_2,\antifund_4) & \quad Y_{34} : (\fund_3,\antifund_4) &
\quad Z_{23} : (\fund_2,\antifund_3) \nonumber\\
X_{42} : (\fund_4,\antifund_2) & \quad Y_{43} : (\fund_4,\antifund_3) &
\quad Z_{32} : (\fund_3,\antifund_2)
\label{specz2z2}
\eeqa
}
and a superpotential
\beqa
W & = \Tr \;[& X_{13} Y_{34} Z_{41} - X_{13} Z_{32} Y_{21} +
X_{31} Y_{12} Z_{23} - X_{31} Z_{14} Y_{43} \nonumber \\
 & & + X_{24} Y_{43} Z_{32} - X_{24} Z_{41} Y_{12} + X_{42} Y_{21} Z_{14}
- X_{42} Z_{23} Y_{34} \;]\;
\label{supz2z2}
\eeqa

We would like to interpret the effect of the blow-ups in this field theory. It
is well-known \cite{dm,dgm} that blow-up modes couple to the field theory as
Fayet-Iliopoulos (FI) terms, which trigger a Higgs breaking of the gauge group
when turned on (when the disappearance of the $U(1)$'s is taken into account,
these modes correspond to baryonic expectation values). The framework to
compute the precise mapping between the resolutions of the singularity and the
Higgsing in the field theory has been provided in \cite{dgm}, and analyzed in
\cite{flops} in the case of the $\IZ_2\times \IZ_2$ orbifold. It is based on
regarding the threefold singularity as the moduli space of the D3-brane field
theory, and on computing this moduli space as a function of the
Fayet-Iliopoulos terms. Below we review this computation for the relevant
resolutions of the  \z2z2 orbifold. Readers interested in the result rather
than in its detailed derivation are adviced to skip the discussion until
subsection II.

\medskip

\subsubsection{Construction of the moduli space}

In what follows we concentrate on the $U(1)^4$ gauge theory of {\rm one}
D3-brane probe, and construct its moduli space as a function of the
corresponding FI coefficients $\zeta_i$. It is clear that the twelve fields
$X_{ij}$, $Y_{ij}$, $Z_{ij}$ (collectively denoted $r_a$, $a=1,\ldots,12$ in
what follows) cannot acquire arbitrary independent vevs. Rather, the F-term
equations allow to express them in terms of the vevs of just six fields,
for instance $X_{13}$, $X_{24}$, $Y_{21}$, $Y_{34}$, $Z_{14}$, $Z_{32}$
(subsequently denoted by ${\bar r}_{\bar a}$, ${\bar a}=1,\ldots, 6$). These
relations can be written
\beq
r_a = \prod_{{\bar b}=1}^6 {\bar r}_{\bar b}^{m_{a{\bar b}}}
\label{xr}
\eeq
where the entries $m_{{\bar b}a}$ (the transpose of the matrix
$M=(m_{a{\bar b}})$) are given by
{\footnotesize
\beqa
\begin{array}{ccccccccccccc}
 & X_{13} & X_{24} & X_{31} & X_{42} & Y_{12} & Y_{21} & Y_{34} & Y_{43}
& Z_{14} & Z_{23} & Z_{32} & Z_{41} \\
X_{13} & 1 & 0 & 0 & 1 & 1 & 0 & 0 & 1 & 0 & 0 & 0 & 0 \\
X_{24} & 0 & 1 & 1 & 0 & -1& 0 & 0 & -1& 0 & 0 & 0 & 0 \\
Y_{21} & 0 & 0 & 0 & 0 & 0 & 1 & 0 & 1 & 0 & 1 & 0 & 1 \\
Y_{34} & 0 & 0 & 0 & 0 & 1 & 0 & 1 & 0 & 0 & -1& 0 & -1\\
Z_{14} & 0 & 0 & -1& -1& 0 & 0 & 0 & 0 & 1 & 1 & 0 & 0 \\
Z_{32} & 0 & 0 & 1 & 1 & 0 & 0 & 0 & 0 & 0 & 0 & 1 & 1
\end{array}
\eeqa
}
Thus, we have $X_{31}=X_{24} Z_{32}/Z_{14}$ (as required from the equation of
motion for $Y_{43}$), $X_{42}=X_{13}Z_{32}/Z_{14}$, and so on.

We would like to find a set of variables $p_{\alpha}$, whose vevs are not
restricted by any F-term equations, and such that the variables $r_a$ can
be expressed as products of $p_{\alpha}$'s in a way consistent with the
relations (\ref{xr}). Namely, we look for relations ${\bar r}_{\bar a}=
\prod_\alpha p_{\alpha}^{t_{{\bar a}\alpha}}$ such that in the
resulting equation
\beqa
r_a = \prod_{\alpha} p_{\alpha}^{\Sigma_{\bar b} m_{a{\bar b}}
t_{{\bar b}\alpha}}
\label{xp}
\eeqa
only positive powers of the $p_{\alpha}$'s appear, $\sum_{\ov b} m_{a{\ov b}}
t_{{\ov b}\alpha} \geq 0$ \footnote{In more toric language, the column vectors
of the matrix $T=(t_{{\ov b}\alpha})$ form a basis of the dual of the cone
spanned by the row vectors of $M=(m_{a{\ov b}})$.}. The matrix
$T=(t_{{\ov b}\alpha})$ can be computed to be
{\footnotesize
\beqa
T= {\pmatrix{
1 & 1 & 0 & 0 & 0 & 0 & 0 & 0 & 1 \cr
0 & 1 & 1 & 0 & 0 & 0 & 0 & 0 & 1 \cr
0 & 0 & 1 & 1 & 1 & 0 & 0 & 0 & 0 \cr
0 & 0 & 1 & 0 & 1 & 1 & 0 & 0 & 0 \cr
0 & 0 & 0 & 0 & 0 & 1 & 1 & 0 & 1 \cr
0 & 0 & 0 & 0 & 0 & 1 & 1 & 1 & 0 \cr
}}
\eeqa
}
so $\alpha$ runs from $1$ to $9$. This matrix implies the relations (\ref{xp})
are, explicitly,
\beqa
& X_{13}=p_1 p_2 p_9 \quad ;\quad Y_{12}=p_1p_5p_6 \quad ;\quad
Z_{14}=p_6p_7p_9 \nonumber\\
& X_{24}=p_2 p_3 p_9 \quad ;\quad Y_{21}=p_3p_4p_5 \quad ;\quad
Z_{23}=p_4p_7p_9 \nonumber\\
& X_{31}=p_2 p_3 p_8 \quad ;\quad Y_{34}=p_3p_5p_6 \quad ;\quad
Z_{32}=p_6p_7p_8 \nonumber\\
& X_{42}=p_1 p_2 p_8 \quad ;\quad Y_{43}=p_1p_4p_5 \quad ;\quad
Z_{41}=p_4p_7p_8
\label{xpexplicit}
\eeqa
However, this parametrization has redundancies, since different assignments
of vevs for the $p_{\alpha}$ may lead to the same vevs for the
${\bar r}_{\bar a}$ (and consequently for the $r_a$). To eliminate this
redundancy we must mod out by a set of $\IC^*$ actions on the $p_\alpha$'s
such that they leave the ${\ov r}_{\ov a}$'s invariant. If we denote by
$q_{n\alpha}$ the weight of $p_{\alpha}$ under the $n^{th}$ $\IC^*$
transformation, the invariance of ${\ov r}_{\ov a}$ is ensured if $\sum_\alpha
t_{{\ov a}\alpha} q_{n\alpha}=0$, that is $TQ^T=0$ in matrix notation.
Equivalently (see \cite{wagm} for further details about the equivalence of the
`holomorphic' and `symplectic' versions of this quotient), this can be
described as introducing a set of $U(1)$ gauge symmetries (with zero FI
terms) for the $p_{\alpha}$'s with charge assignments $q_{n{\alpha}}$, such
that the ${\ov r}_{\ov a}$ are gauge invariant composite operators (so they
parametrize the directions which are D-flat with respect to these $U(1)$
symmetries). A possible matrix Q is given by
{\footnotesize
\beqa
Q=\pmatrix{
0 & 0 & 0 & 1 & -1& 1 & -1& 0 & 0 \cr
0 & 1 & 0 & 0 & 0 & 0 & 1 & -1& -1 \cr
1 &-1 & 1 & 0 & -1& 0 & 0 & 0 & 0 \cr }
\eeqa
}
This provides a symplectic quotient description of the F-flat direction of
our original theory (\ref{specz2z2}). At this point we should recall that
the vevs for the $r_a$'s in the original theory were also constrained by
D-flatness conditions. These constraints can be imposed at the level of the
$p_{\alpha}$ if we can find an assignment of charges $q_{i{\alpha}}$ for
$p_{\alpha}$, such that it reproduces the charge $v_{i{\bar a}}$ of
${\bar r}_{\bar a}$ under the $U(1)_i$, $i=1,2,3$ in (\ref{specz2z2})
{\small
\beqa
V=(v_{i\alpha}) =\pmatrix{
1 & 0 & -1& 0 & 1 & 0 \cr
0 & 1 & 1 & 0 & 0 & -1\cr
-1& 0 & 0 & 1 & 0 & 1 \cr}
\eeqa
}
One such assignment is $q_{i\alpha}=\sum_{{\ov b}=1}^6 v_{i{\ov b}}
u_{{\ov b}\alpha}$, with $U=(u_{{\ov b}\alpha})$ a matrix satisfying $TU^T=1$
(so that ${\bar r}_{\bar a}$ has the correct charge $\sum_{\alpha}
q_{i\alpha}t_{{\bar a}\alpha}=v_{i{\bar a}}$). One possible choice for $U$ is
{\small
\beqa
U=\pmatrix{
1 & 0 & 0 & 0 & 0 & 0 & 0 & 0 & 0 \cr
-1& 1 & 0 & 0 & 0 & 0 & 0 & 0 & 0 \cr
0 & 0 & 0 & 1 & 0 & 0 & 0 & 0 & 0 \cr
0 & 0 & 0 & 0 & 0 & 1 & -1& 0 & 0 \cr
0 & -1& 0 & 0 & 0 & 0 & 0 & 0 & 1 \cr
0 & 0 & 0 & 0 & 0 & 0 & 0 & 1 & 0 \cr
}
\eeqa
}
The moduli space of D- and F-flat directions in the original theory
(\ref{specz2z2}) is obtained by the quotienting the space spanned by the
$p_{\alpha}$ by the combined action of $U(1)$ symmetries under which
$p_{\alpha}$ has charges $q_{n{\alpha}}$, $q_{i\alpha}$. The complete charge
matrix (concatenation of $Q$ and $VU$) is
{\footnotesize
\beqa
{\tilde Q}=\pmatrix{
0 & 0 & 0 & 1 & -1& 1 & -1& 0 & 0 &  0\cr
0 & 1 & 0 & 0 & 0 & 0 & 1 & -1& -1&  0 \cr
1 & -1& 1 & 0 & -1& 0 & 0 & 0 & 0 &  0 \cr
1 & -1& 0 & -1& 0 & 0 & 0 & 0 & 1 & \zeta_1\cr
-1& 1 & 0 & 1 & 0 & 0 & 0 & -1& 0 & \zeta_2\cr
-1& 0 & 0 & 0 & 0 & 1 & -1& 1 & 0 & \zeta_3\cr }
\label{fincharg}
\eeqa
}
where we have allowed arbitrary FI terms (indicated in the last column) for the
$U(1)$'s of the original theory. The FI for $U(1)_4$, $\zeta_4$, is not
an independent parameter, since the existence of supersymmetric vacua
imposes $\sum_{i=1}^4=0$.

\begin{figure}
\centering
\epsfxsize=3.5in
\hspace*{0in}\vspace*{.2in}
\epsffile{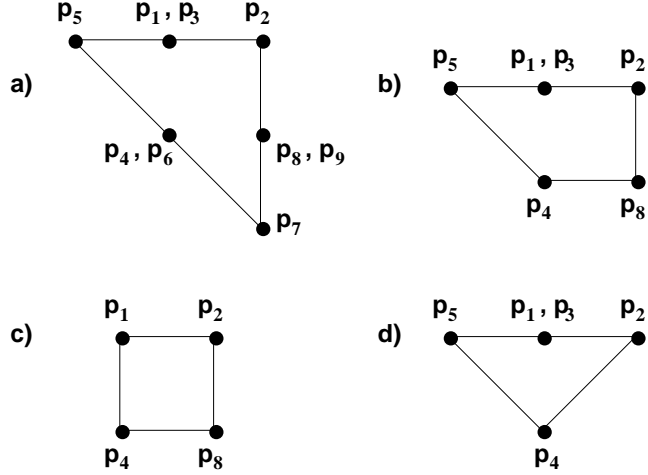}
\caption{\small The toric diagrams corresponding to {\bf a)} the \z2z2
orbifold, $xyz=w^2$, {\bf b)} the suspended pinchpoint singularity,
$xy=zw^2$, {\bf c)} the conifold, $xy=zw$, and {\bf d)} the
$\IZ_2$ orbifold, $xy=w^2$.}
\label{fig:toric}
\end{figure}

The symplectic quotient description provides automatically a toric description
of the moduli space. The toric data (vectors of the fan) are given by the
transpose of the kernel of ${\tilde Q}$. When $\zeta_i=0$, the vector defining
the toric data are given by the columns of \vspace*{-.5cm}
{\small
\beqa
{\tilde T}=\pmatrix{
0 & 1 & 0 & 0 & -1& 0 & 1 & 1 & 1 \cr
1 & 1 & 1 & 0 & 1 & 0 & -1& 0 & 0 \cr
1 & 1 & 1 & 1 & 1 & 1 & 1 & 1 & 1 \cr
}
\eeqa
}
The fact that all vector endpoints lie on a plane ensures the threefold is
Calabi-Yau. The polygon these points define, shown in figure
\ref{fig:toric}a, can be seen to correspond to \z2z2. This can be
understood directly by defining the invariant variables
\beqa
\begin{array}{lllll}
x = p_1p_2^2 p_3 p_8 p_9  & (=X_{13}X_{31}) & \quad &
y = p_1 p_3 p_4 p_5^2 p_6 & (=Y_{34}Y_{43}) \\
z = p_4 p_6 p_7^2 p_8 p_9 & (=Z_{14}Z_{41}) & \quad &
w=p_1 p_2 p_3 p_4 p_5 p_6 p_7 p_8 p_9 &(=X_{13}Y_{34}Z_{41})
\end{array}
\label{vars}
\eeqa
which are related as $xyz=w^2$. From their expression in terms of the
original fields (\ref{specz2z2}), given in parentheses, we note
that these variables indeed span the moduli space, regarded as the space of
gauge invariant operators of the field theory (\ref{specz2z2}) modulo the
equations of motion.

We are now ready to consider the partial resolutions that arise when we
consider non-vanishing FI terms. We will not perform an exhaustive
exploration of the parameter space, but rather present one example of each
possible blow-up. This will illustrate the type of argument we will need
for our future purposes.

{\bf i)} Let us consider a single blow-up of this space. This can be done
by turning on one FI term, so we consider $\zeta_1\gg 0$, $\zeta_2=\zeta_3=0$.
A solution to the D-term equations for the charge matrix (\ref{fincharg})
is provided by $|p_6|^2=|p_7|^2=|p_9|^2=\zeta_1$. These vevs break three
of the $U(1)$ gauge symmetries. The fields $p_6$, $p_7$, $p_9$ disappear, and
the remaining fields $p_1$, $p_2$, $p_3$, $p_4$, $p_5$, $p_8$ can still take
vevs which are constrained only by the D-terms of the unbroken $U(1)$'s.
The corresponding charge matrix and toric data are
{\small
\beqa
{\tilde Q}_{SPP} = \pmatrix{
1 & -1& 1 & 0 & -1& 0 & 0 \cr
-1& 1 & 0 & 1 & 0 & -1 & \zeta_2 \cr
-1& 0 & 0 & -1& 1 & 1 & \zeta_3 \cr
} \quad ; \quad
{\tilde T}_{SPP}= \pmatrix{
0 & 1 & 0 & 0 & -1& 1 \cr
1 & 1 & 1 & 0 & 1 & 0 \cr
1 & 1 & 1 & 1 & 1 & 1 \cr
}
\eeqa
}
The toric diagram, shown in figure \ref{fig:toric}b, is known to describe the
suspended pinch point singularity. We can see this directly, by noticing
that, modulo constant coefficients, the variables (\ref{vars}) read
\beqa
x=p_1 p_2^2 p_3 p_8 \;\; ,\;\; y= p_1 p_3 p_4 p_5^2 \;\; ,\;\;
z=p_4 p_8 \;\; ,\;\; w=p_1 p_2 p_3 p_4 p_5 p_8
\label{varspp}
\eeqa
However, there exists a new variable $w'=p_2 p_2 p_3 p_5$, invariant under
all the surviving $U(1)$'s. Notice that $w=w'z$, so $w'$ parametrizes a
blown-up $\IP_1$. In terms of the basic invariants $x$, $y$, $z$, $w'$,
the blown-up space is described by $xy=zw'^2$.

Before continuing with further blow-ups, we would like to identify the
above vev in terms of the original fields (\ref{specz2z2}). Recalling the
relations (\ref{xpexplicit}), we see that a vev for $p_6$, $p_7$, $p_9$
corresponds to a vev for the field $Z_{14}$.

\medskip

{\bf ii)} Let us perform a further blow up, by allowing a non-vanishing
$\zeta_3$. Thus we consider $\zeta_1\gg 0$, $\zeta_3\gg 0$, $\zeta_2=0$.
The D-term equations are solved by $|p_3|^2=|p_5|^2=\zeta_3$, $|p_6|^2= \zeta_1
+\zeta_3$, $|p_7|^2=|p_9|^2=\zeta_1$. The remaining fields $p_1$, $p_2$, $p_4$,
$p_8$ have the following charge matrix and toric data
{\small
\beqa
{\tilde Q}_{\rm conif.} =\pmatrix{
-1 & 1 & 1 & -1 & \zeta_2\cr
} \quad ; \quad
{\tilde T}_{\rm conif.}=\pmatrix{
0 & 1 & 0 & 1 \cr
1 & 1 & 0 & 0 \cr
1 & 1 & 1 & 1 \cr
}
\eeqa
}
The toric diagram is shown in figure \ref{fig:toric}c, and corresponds to
the conifold singularity. This can be seen from the invariant variables
(\ref{varspp}), which after the Higgsing read
\beqa
x=p_1p_2^2 p_8 \;\; ,\;\; y=p_1p_4 \;\; ,\;\; z=p_4p_8 \;\; ,\;\;
w'=p_1p_2
\eeqa
There is a new invariant, $x'=p_2p_8$, such that $x=x'w'$. The remaining
singularity is $x'y=zw'$, a conifold. Using (\ref{xpexplicit}) we see that
the blow-up from \z2z2 to the conifold corresponds to a vev for the fields
$Z_{14}$, $Y_{34}$.

\medskip

{\bf iii)} For completeness, let us also discuss the blow-up of the SPP
to the $\IZ_2$ orbifold singularity. In order to do that, we consider
$\zeta_1\gg 0$, $\zeta_2\ll 0$, $\zeta_3\gg 0$, $\zeta_3+\zeta_2=0$. The
constraints are solved by the vevs $|p_6|^2=|p_7|^2=\zeta_1-\zeta_2$,
$|p_8|^2=-\zeta_2$, $|p_9|^2=\zeta_1$. The remaining fields $p_1$, $p_2$,
$p_3$, $p_4$, $p_5$ have the following charge matrix and toric data
{\small
\beqa
{\tilde Q}_{\IZ_2} =\pmatrix{
1 & -1 & 1 & 0 & -1 & 0 \cr
-2 & 1 & 0 & 0 & 1 & \zeta_2+\zeta_3 \cr
} \quad ; \quad
{\tilde T}_{\IZ_2}=\pmatrix{
0 & 1 & 0 & 0 & -1 \cr
1 & 1 & 1 & 0 & 1 \cr
1 & 1 & 1 & 1 & 1 \cr
}
\eeqa
}
The toric diagram is depicted in figure \ref{fig:toric}d. The blown-up space
corresponds to $\IC^2/\IZ_2\times \IC$, as can be seen by looking at the
invariant variables (\ref{varspp}), which now read
\beqa
x=p_1p_2^2 p_3 \;\; ,\;\; y= p_1p_3p_5^2 \;\; ,\;\; z=p_4 \;\; ,\;\;
w'=p_1p_2p_3p_5
\eeqa
The new invariant is $y'=p_1p_3p_5^2=y/z$. The remaining singularity is
$xy'=w'^2$. Finally, let us mention that the blow-up from \z2z2 to the $\IZ_2$
orbifold can be seen from (\ref{xpexplicit}) to correspond to vevs for
$Z_{14}$, $Z_{32}$.

\medskip

This concludes our review of the toric description of \z2z2 and its
blow-ups, so we turn to the construction of the field theories obtained
after following the Higgs branches we have just mentioned. In what follows
we go back to the case of more general ranks for the gauge groups in
(\ref{specz2z2}), and moreover take into account the freezing of the
$U(1)$ factors. Thus the Higgs breakings are interpreted as baryonic
branches.

\medskip

\subsubsection{Back to the field theories}

Now we are in good shape to interpret the blowing-ups of \z2z2 from the
field theory viewpoint. This allows us to construct the field theories at
non-orbifold singularities.

{\bf i)} Let us construct the field theory of D3-branes at the SPP singularity
$xy=zw^2$. As determined above, it is obtained from (\ref{specz2z2}) by
turning on a suitable blow-up mode, which corresponds to giving a diagonal vev
to $Z_{14}$. This flat direction exists iff $n_1=n_4$ (subsequently denoted by
$n$) and implies the breaking to the diagonal subgroup $SU(n)_1\times SU(n)_4
\to SU(n)_{(14)}$. The field $Z_{14}$ is swallowed by the Higgs mechanism, and
the superpotential (\ref{supz2z2}) makes the fields $X_{31}$, $Y_{43}$,
$X_{42}$, $Y_{21}$ massive. The surviving light fields are
{\small
\beqa
\begin{array}{cccc}
& SU(n)_{(14)} & SU(n_2) & SU(n_3)\\
Y_{12} & \fund & \antifund &  \\
X_{24} & \antifund & \fund &  \\
Z_{23} & & \fund & \antifund   \\
Z_{32} & & \antifund & \fund   \\
Y_{34} & \antifund & & \fund \\
X_{13} & \fund & & \antifund \\
Z_{41} & {\rm Adj.} & &
\label{specspptwo}
\end{array}
\eeqa
}
Integrating out the massive fields using their equations of motion, we obtain
the superpotential
\beqa
W=\Tr (X_{24} Y_{12} Z_{23} Z_{32}- Z_{32} Z_{23} Y_{34} Z_{13} + X_{13}
Y_{34} Z_{41} - X_{24} Z_{41} Y_{12})
\label{supspptwo}
\eeqa
This field theory agrees with (\ref{specspp}), (\ref{supspp}) by an
obvious relabeling of fields.

{\bf ii)} The SPP singularity can be further blown-up to a conifold. This
allows to construct the field theory of D3-branes at the conifold by taking a
baryonic branch from the theory above. As studied before, the suitable
Higgsing is achieved by giving a diagonal vev to $Y_{34}$. This vev is
possible when $n=n_3$, and triggers the breaking of $SU(n)_{(14)}\times
SU(n)_3 \to SU(n)_{(134)}$. The field $Y_{34}$ is swallowed, and $X_{13}$,
$Z_{41}$ become massive. The remaining light fields are
{\small
\beqa
\begin{array}{ccc}
& SU(n)_{(134)} & SU(n_2)  \\
Y_{12}, Z_{32} & \fund & \antifund \\
X_{24}, Z_{23} & \antifund & \fund
\end{array}
\label{coniagain}
\eeqa
}
Integrating out the massive fields, the superpotential is
\beqa
W=\Tr (Y_{12}Z_{23}Z_{32}X_{24}-Y_{12}X_{24}Z_{32}Z_{23})
\eeqa
This field theory agrees with (\ref{speconi}), (\ref{supconi}). Notice that
the Higgsing we have just discussed has a nice interpretation in the IIA
picture, where it corresponds to removing one NS$'$ brane from the
configuration in figure \ref{fig:conspp}b to recover figure
\ref{fig:conspp}a.

{\bf iii)} For completeness, we also discuss the field theory interpretation
of the blow-up of the SPP to the $\IZ_2$ orbifold. As discussed above, it
corresponds to Higgsing the field theory (\ref{specspptwo}), (\ref{supspptwo})
with a diagonal vev for $Z_{32}$. This is possible when $n_2=n_3$ (denoted by
$m$ in what follows), and triggers the breaking $SU(m)_2\times SU(m)_3 \to
SU(m)_{(23)}$. The field $Z_{32}$ disappears, but no field becomes massive.
The surviving light fields are
{\small
\beqa
\begin{array}{ccc}
         & SU(n)_{(14)} & SU(m)_{(23)}\\
Y_{12}, X_{13} & \fund & \antifund \\
X_{24}, Y_{34} & \antifund & \fund \\
Z_{41} & {\rm Adj.} & \\
Z_{23} & & {\rm Adj.}
\end{array}
\eeqa
}
and their superpotential is
\beq
W=\Tr (X_{13}Y_{34}Z_{41}-X_{24} Z_{41}Y_{12}+ Y_{12}Z_{23}X_{24}-
Y_{34}X_{13}Z_{23})
\eeq
which agrees with the field theory of D3-branes at a $\IZ_2$ singularity
\cite{dm}. This Higgs breaking can be interpreted in the IIA side as the
removal of the NS-brane from the configuration shown in figure
\ref{fig:conspp}b.

\section{Introduction of orientifold planes in the IIA side}

In this section we start the study of orientifolded models by describing the
different IIA brane configurations that can be obtained introducing O6-planes
or O6$'$-planes in the $\NN=1$ elliptic models described in Section~1. The
type IIA construction of more general models is straightforward
\cite{uraconi,ehn}. However, the computation of the corresponding type IIB
T-duals would be extremely involved, so we restrict to the simplest examples.

\subsection{Models with one NS-brane and two NS$'$-branes}

\begin{figure}
\centering
\epsfxsize=5.5in
\hspace*{0in}\vspace*{.2in}
\epsffile{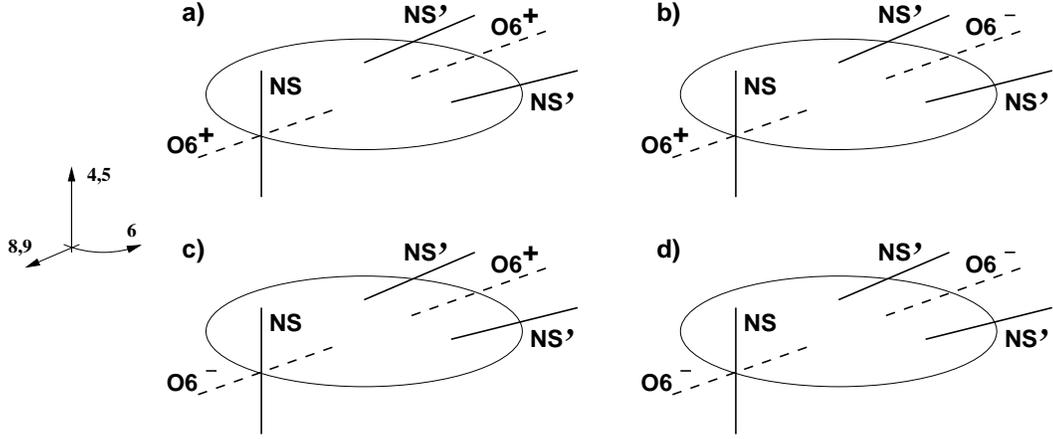}
\caption{\small Non-chiral orientifolds of the IIA brane configuration
T-dual to the suspended pinch point singularity.
}
\label{fig:orspp1}
\end{figure}

Let us first consider the IIA configuration with one NS-brane and two
NS$'$-branes. The different brane configurations that arise when we introduce
O6-planes (along 0123789) are shown in Figure~\ref{fig:orspp1}. The two
NS$'$-branes are necessarily related to each other by the orientifold
projection, while the NS-brane is mapped to itself and is therefore stuck at
one of the O6-planes. As usual, different field theories arise from the
different choices of the O6-plane charges. The matter content and interactions
are obtained using the rules in \cite{landlop,bhkl}. For instance, the
configuration with two negatively charged O6-planes gives the following field
theory content
{\small
\beqa
\begin{array}{ccc}
& USp(n_1) & SU(n_2) \\
Q & \fund & \antifund \\
{\tilde Q} & \fund & \fund \\
A_0 & \Yasymm & \\
A_{1} & & \Yasymm \\
{\tilde A}_1 & & \bYasymm
\end{array}
\eeqa
}
There is a superpotential given by
\beqa
W= A_0 Q {\tilde Q}-{\tilde Q} Q A_1 {\tilde A}_1
\eeqa

The field theory for the remaining cases is obtained by obvious replacements
of antisymmetric representations by symmetric ones, and/or symplectic gauge
factors by orthogonal factors. All these theories are non-chiral.

\begin{figure}
\centering
\epsfxsize=5.5in
\hspace*{0in}\vspace*{.2in}
\epsffile{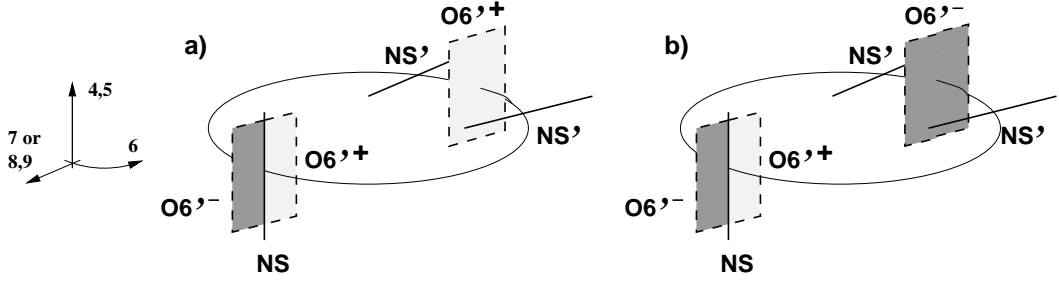}
\caption{\small Chiral orientifolds of the IIA brane configuration T-dual
to the suspended pinch point singularity. Note that O6$'$-planes extend
along 7 (but not 89) and NS$'$-branes extend along 89 (but not 7).
}
\label{fig:orspp2}
\end{figure}

Another possibility is to introduce O6$'$-planes, as illustrated in figure
\ref{fig:orspp2}. As before, the two NS$'$-branes are mapped to each other
under the orientifold projection, while the NS-brane is invariant. However,
in this case the NS-brane divides the O6$'$-plane in two halves, which
must have different RR charge \cite{ejs}. The model requires eight half
D6$'$-brane to conserve RR charge, and yields a chiral spectrum. We refer
to this sector as the `fork' configuration\cite{lll,egkt,bhkl}. There are
two possible field theories, which differ in the choice of orientifold
charge. The field theory corresponding to the configuration with one fork
and one O6$'\,^-$ has the following gauge group and matter content
{\small
\beqa
\begin{array}{cccc}
& USp(n_1) & SU(n_2) & \\
Q & \fund & \antifund &\\
{\tilde Q} & \fund & \fund &\\
S_1 & \Ysymm  &\\
{\tilde A}_{2} & & \bYasymm & \\
S_2 & & {\Ysymm} & \\
{\tilde T}_2^{a} & & \antifund & a=1,\ldots,8\\
\end{array}
\label{specfine}
\eeqa
}
The superpotential is given by
\beqa
W= S_1 Q{\tilde Q}-{\tilde Q} Q {\tilde A}_2 S_2 + S_2 {\tilde T}_2
{\tilde T}_2
\eeqa
The field theory for the configuration with the O6$'\,^+$-plane has a
similar structure.

\medskip

\subsection{Models with one NS-brane and one NS$'$-brane}

\begin{figure}
\centering
\epsfxsize=5.5in
\hspace*{0in}\vspace*{.2in}
\epsffile{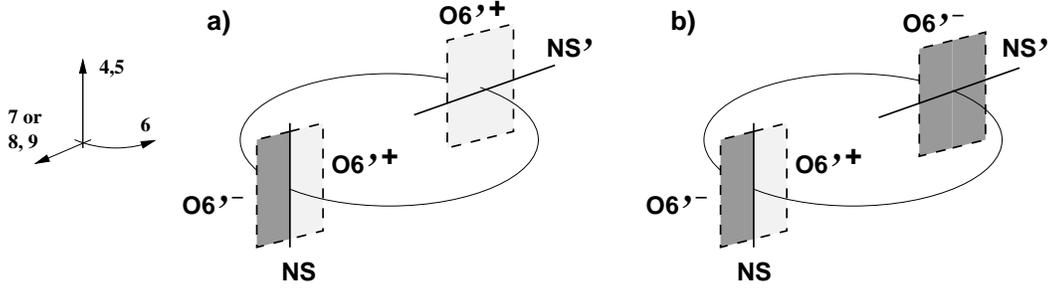}
\caption{\small The introduction of O6$'$-planes in the IIA brane
configuration T-dual to the conifold singularity. The picture appears
confusing since we have tried to show too many dimensions in it. The only
point to keep in mind is that a NS$'$-brane does not split an O6$'$-plane
in two halves.
}
\label{fig:orconi1}
\end{figure}

Let us turn to the IIA configurations with one NS-brane and one NS$'$-brane.
In this case the most obvious possibility is that each fivebrane is mapped to
itself, so each is stuck at one orientifold plane. The theories obtained
by introducing O6$'$-planes and O6-planes are  equivalent, and for
concreteness we discuss the configuration with O6$'$-planes, depicted
in figure \ref{fig:orconi1}. One of the O6$'$-planes is split in halves by
the NS-brane, so the configuration contains one fork. There are two field
theories, depending on the sign of the O6$'$-plane intersected by the
NS$'$-brane. Choosing for instance the configuration in figure
\ref{fig:orconi1}b, the resulting field theory is
{\small
\beqa
\begin{array}{ccc}
& SU(n_1) & \\
A & \Yasymm  & \\
{\tilde A} & \bYasymm & \\
{\tilde A}' & \bYasymm & \\
S' & {\Ysymm} & \\
{\tilde T}'^{a} & \antifund & a=1,\ldots,8\\
\end{array}
\label{sporcon}
\eeqa
}
There is a superpotential given by
\beqa
W= A{\tilde A}{\tilde A}' S' - S'{\tilde T}' {\tilde T}'
\eeqa

A less obvious configuration is possible, in which we consider
NS-fivebranes rotated $\pm 45^0$ in the 45-89 plane. In such case, an
O6$'$-plane can map the fivebranes to one another, so they must be located
at $\IZ_2$ symmetric positions in $x^6$ (note that the configuration with
O6-planes provides an equivalent model). We will not consider this case in
the present paper.

\section{T-dual models I: Orientifolds of $xy=zw^2$}

Our aim in the rest of the paper is to construct the type IIB T-duals of
these configurations. They are expected to be given by suitable
orientifolds of the singularities $xy=zw^2$ and $xy=zw$. However, it is
not obvious how to check this directly. Even though the T-duality allows
to guess the action of the orientifold projection on these spaces, the
usual techniques to compute the spectrum and to check the cancellation of
tadpoles \cite{sag1,dlp,horava,bs,gp,gjdp} are only valid for models with 
a solvable worldsheet theory. In this section we show how these 
difficulties can be overcome,  using an indirect approach based on the 
techniques reviewed in section 2.2.

More concretely, we will construct configurations of D3-branes at
orientifolds of \z2z2, and perform partial resolutions of the singular
geometry to obtain orientifolds of the SPP and conifold singularities. As in
section~2.2, these blow-ups correspond to baryonic branches in the D3-brane
field theory. Since the orientifold models are obtained by imposing $\IZ_2$
identifications on the fields in the orbifold theory, any baryonic branch
in the orientifold can be regarded as inherited from baryonic branches in
the orbifold model. Hence the the map between blow-up modes and baryonic
Higgsings we found in section 2.2 will be useful for our analysis. The
reverse implication, though, does not work, and certain blow-ups in the
orbifold case may be absent, `frozen', in the orientifold models. We will
find an example of this in Section~5.

\subsection{Some orientifolds of \z2z2}

In this section we construct several orientifolds of \z2z2, explicitly
checking their consistency (algebraic consistency and cancellation of
tadpoles), and computing the field theories arising on D3-brane probes. The
list below is {\em not} intended to be a complete classification of such
models. Rather, it provides us with a rich enough starting point for
recovering, upon blowing-up, the field theories obtained in section 3.1
from the IIA perspective. Since the computations of orientifold spectra
are rather standard, we leave most of the details out of the discussion, and
merely describe the models. Cancellation of tadpoles is discussed in more
detail in appendix A. We consider four different models, referred to as
A1, A2, B and C.

{\bf Models A}

We consider the following structure for the orientifold group
\beqa
(1+\theta+\omega+\theta\omega)(1+\Omega')
\eeqa
where $\Omega'=\Omega(-1)^{F_L}R_1R_2R_3$. We also take the D3-brane
Chan-Paton matrices
\beqa
& \gamma_{\theta,3}  =  \diag (i 1_{n_1},i 1_{n_2},-i 1_{n_2}, -i 1_{n_1})
\nonumber \\
& \gamma_{\omega,3} =  \diag (i 1_{n_1},-i 1_{n_2},i 1_{n_2}, -i 1_{n_1})
\nonumber \\
& \gamma_{\Omega',3}  = {\small \left( \begin{array}{cccc}
  & & & 1_{n_1} \\
  & & 1_{n_2} & \\
  & 1_{n_2}& &  \\
  1_{n_1}& & &  \\
\end{array}  \right)} \quad  {\rm or} \quad
\gamma_{\Omega',3}  = {\small \left( \begin{array}{cccc}
  & & & 1_{n_1} \\
  & & 1_{n_2} & \\
  & -1_{n_2}& &  \\
  -1_{n_1}& & &  \\
\end{array}  \right) } \nonumber\\
\eeqa
where the choice of $\gamma_{\Omega'}$ defines the choice of $SO$ or $Sp$
projection on the D3-branes. We refer to these models as A1 and A2,
respectively. The spectra obtained from the orientifold projections are
provided below, in eqs. (\ref{specorz2z2}), (\ref{twotens}). As discussed
in appendix A, this model is consistent without the addition of D7-branes.

{\bf Model B}

Further models can be obtained by changing the orientifold group. Let us
consider
\beqa
(1+\theta+\omega+\theta\omega)(1+\beta\Omega')
\eeqa
where $\beta:\,(z_1,z_2,z_3) \to (iz_1,-iz_2,z_3)$, that is $\beta^2 = \theta
\omega$. A suitable Chan-Paton embedding is defined by \vspace*{-.5cm}
\beqa
\begin{array}{c}
\gamma_{\theta,3}  =  \diag (i 1_{n_1},i 1_{n_2},-i 1_{n_2}, -i 1_{n_1}) \\
\gamma_{\omega,3}  =  \diag (i 1_{n_1},-i 1_{n_2},i 1_{n_2}, -i 1_{n_1})
\end{array} &
\gamma_{\beta\Omega',3}  = {\small \left(
\begin{array}{cccc}
  & & & i \varepsilon_{n_1} \\
  & & 1_{n_2} & \\
  & - 1_{n_2}& &  \\
  -i\varepsilon_{n_1}& & &  \\
\end{array}  \right)}
\eeqa
In this case, the matrix $\gamma_{\beta\Omega',3}$ is rather unique, and
we obtain only one field theory. Its spectrum is given below, in eqs.
(\ref{specorz2z2}), (\ref{twotens}). As shown in appendix, this models
does not require D7-branes for consistency.

{\bf Model C}

Our last model is constructed using the orientifold group
\beqa
(1+\theta+\omega+\theta\omega)(1+\alpha\Omega')
\label{eq:oc}
\eeqa
where $\alpha:\,(z_1,z_2,z_3) \to (iz_1,z_2,-iz_3)$, that is $\alpha^2
= \theta$). Our choice of Chan-Paton matrices is \vspace{-.5cm}
\beqa
\begin{array}{c}
\gamma_{\theta,3}  =  \diag (i 1_{n_1},i 1_{n_2},-i 1_{n_2}, -i 1_{n_1}) \\
\gamma_{\omega,3}  =  \diag (i 1_{n_1},-i 1_{n_2},i 1_{n_2}, -i 1_{n_1})
\end{array} &
\gamma_{\alpha\Omega',3}  = {\small \left( \begin{array}{cccc}
  & & & e^{\frac{\pi i}{4}} 1_{n_1} \\
  & & e^{\frac{\pi i}{4}}1_{n_2} & \\
  & e^{\frac{3\pi i}{4}} 1_{n_2}& &  \\
  e^{\frac{3\pi i}{4}} 1_{n_1}& & &  \\
\end{array}  \right)} \quad\quad
\eeqa
This model contain non-vanishing Klein bottle tadpoles. As shown in appendix
A they can be cancelled by introducing a set of D7$_3$ branes with
Chan-Paton factors:
\beqa
\begin{array}{c}
\gamma_{\theta,7_3}  =  \diag (1_{m_1}, -1_{m_2},- 1_{m_3}, 1_{m_4}) \\
\gamma_{\omega}  =  \diag (1_{m_1}, -1_{m_2}, 1_{m_3}, -1_{m-4})
\end{array} &
\gamma_{\alpha\Omega'}  = {\small \left( \begin{array}{cccc}
\varepsilon_{m_1}  & & & \\
  & 1_{m_2}& & \\
  & & 1_{m_3} &  \\
  & & & \varepsilon_{m_4} \\
\end{array}  \right)}
\label{eq:occ}
\eeqa
subject to the tadpole cancellation constraint
\beqa
m_2-m_1=8 \quad ;\quad m_3-m_4=8
\label{tadmodelc}
\eeqa

\medskip

The spectra on the world-volume of the D3-brane probes in all these models
are quite similar. Their basic structure is given by \footnote{Here we
are taken into account the disappearance of the $U(1)$ factors.}
{\small
\beqa
\begin{array}{ccc}
& SU(n_1) & SU(n_2) \\
X_{13} & \fund & \fund \\
X_{31} & \antifund & \antifund \\
Y_{12} & \fund & \antifund \\
Y_{21} & \antifund & \fund \\
Z_{14} & {\rm R} & \\
Z_{41} & {\ov {\rm R}} & \\
Z_{23} &  & {\rm R} \\
Z_{32} &  & {\ov {\rm R}}
\end{array}
\label{specorz2z2}
\eeqa
}
where ${\rm R}$ represents different two-index tensor representations,
which are model-dependent. They are given explicitly in the following table
\beqa
\begin{array}{ccccc}
 & A1 & A2 & B & C \\
Z_{14} & \Yasymm_1 & \Ysymm_1 & \Yasymm_1 & \Ysymm_1 \\
Z_{41} & \bYasymm_1 & {\ov{\Ysymm}}_1 & \bYasymm_1 & \bYasymm_1 \\
Z_{23} & \Yasymm_2 & \Ysymm_2 & \Ysymm_2 & \Ysymm_2 \\
Z_{32} & \bYasymm_2 & {\ov{\Ysymm}}_2 & {\ov{\Ysymm}}_2 & \bYasymm_2
\label{twotens}
\end{array}
\eeqa
The model C has additional fields coming from the 3-7$_3$ sector. For the
minimal choice of Chan-Paton matrices satisfying (\ref{tadmodelc}),
$m_2=m_3=8$, $m_1=m_4=0$, we have eight fields $T^a_1$ in the anti-fundamental
of $SU(n_1)$ and eight fields $U^a_2$ in the anti-fundamental of $SU(n_2)$.
The global $SO(8)^2$ global symmetry acting on these flavours is realized 
on the world-volume of the corresponding D7-branes. The model C is the 
only chiral one.

Comparing the spectra above with the orbifold spectrum (\ref{specz2z2}),
we see the effect of the orientifold projection is to identify the groups
$SU(n_1)\leftrightarrow SU(n_4)$ (in such a way that $\fund_1
\leftrightarrow \antifund_4$) and $SU(n_2)\leftrightarrow SU(n_3)$ (so
that $\fund_2 \leftrightarrow \antifund_3$. This implies the following
$\IZ_2$ identification in the fields of the orbifold theory
\beqa
\begin{array}{cccc}
X_{13} \leftrightarrow X_{24}\; , & Y_{12} \leftrightarrow Y_{34}\; ,
& Z_{14} \leftrightarrow Z_{14}\; , & Z_{23} \leftrightarrow Z_{23}\; , \\
X_{31} \leftrightarrow X_{42}\; , &Y_{21} \leftrightarrow Y_{43}\; ,
& Z_{41} \leftrightarrow Z_{41}\; , & Z_{32} \leftrightarrow Z_{32}\; ,
\end{array}
\label{involution}
\eeqa
The models A1, A2, B, C differ in the introduction of different signs in
this identifications. These are particularly important, since they determine
the symmetry of the two-index tensor representations (\ref{twotens}).
However, the relation above contains the relevant information for many
purposes. For instance, the basic structure of the superpotential in the
orientifold models is obtained from (\ref{supz2z2}) upon imposing the
identification above.

In the following section we study the different resolutions of these
orientifolds. Since the computations are familiar from the orbifold case,
we discuss the details only for the most interesting case, the chiral
theory, model C. Its superpotential is
\beqa
W= \Tr [\; X_{13}^T Z_{41} Y_{12} - X_{13} Z_{32} Y_{21} + X_{31} Y_{12}
Z_{23} + X_{31} Z_{14} Y_{21}^T + Z_{14} U U + Z_{23} T T \; ] \quad
\label{supmodelc}
\eeqa

For future convenience, we also list in the following table the action of the
orientifold projection on \z2z2, when described as $xyz=w^2$ in terms of the
invariant variables $x=z_1^2$, $y=z_2^2$, $z=z_3^2$, $w=z_1z_2z_3$:
\beqa
\begin{array}{ccccc}
{\rm A1, \; A2}: & x\to x\; , & y \to y\; , & z\to z\; , & w\to -w \\
{\rm B}:         & x\to -x\; , & y \to -y\; , & z\to z\; , & w\to -w \\
{\rm C}:         & x\to -x\; , & y \to y\; , & z\to -z\; , & w\to -w \\
\end{array}
\label{orz2z2}
\eeqa

We end this section with a toric comment. Using (\ref{xpexplicit}), the
relations (\ref{involution}) show that the orientifold action can be
implemented at the level of the $p_{\alpha}$'s of section 2.2.1 as the
$\IZ_2$ identification \vspace*{-.5cm}
\beqa
p_1 \leftrightarrow p_3
\label{invop}
\eeqa
In fact, the relation above does not encode the differences between the models
A, B, C. This could be easily accomplished by using matrix-valued fields
$p_{\alpha}$, but we will not need this refinement. For future convenience, we
note that, given the action of the orientifold on the gauge groups of the
orbifold theory (see comment preceding (\ref{involution})), it also imposes
the following relation \footnote{The relation below can also be obtained by
imposing the conditions that (\ref{invop}) is a symmetry of the charge matrix
${\tilde Q}$ (\ref{fincharg}).} on the FI terms of section 2.2.1
\beqa
\zeta_1=-\zeta_4 \;\; ,\;\; \zeta_2=-\zeta_3
\label{invozeta}
\eeqa
where $\zeta_4= -\sum_{i=1}^3 \zeta_i$ is the FI for the last $U(1)_4$.

\subsection{Orientifolds of $xy=zw^2$}

In this section we perform a single blow-up in the orientifolds of \z2z2
constructed in the previous section, in order to reproduce orientifolds of
the suspended pinch point singularity, $xy=zw^2$. We show that D3-brane
probing the different orientifolds of this singularity realize the different
field theories constructed in section 3.1 from the type IIA viewpoint, as
orientifolded $\NN=1$ elliptic models. As in \cite{pru}, these constructions
illustrate how several exotic brane dynamics effects on the IIA side have a
quite standard realization in the type IIB T-dual setup.

The first observation is that, in order to obtain orientifolds of the SPP
variety, the required blow-up must correspond to a vev for one of the fields
$Z_{ij}$, rather than to the fields $X_{ij}$ or $Y_{ij}$. This follows from
the fact that a vev for one of the latter fields in the orientifold model
can be thought to arise from a vev for {\em two} fields, related by
(\ref{involution}), in the orbifold theory. The analysis in section 2.2.1
shows that the associated blow-ups resolve \z2z2 to the $\IZ_2$ orbifold.
Orientifolds of $xy=xw^2$ are therefore only obtained by following baryonic
branches associated to the fields $Z_{ij}$, invariant under
(\ref{involution}). Let us discuss the models obtained starting with the
chiral \z2z2 orientifold, model C.

Consider giving a vev to the symmetric representation $Z_{14}$. From our
experience in section 2.2, we know this corresponds to blowing-up by
introducing a $\IP_1$ parametrized by $w'=w/z$, so the resulting space is an
orientifold of $xy=zw'^2$. For readers acquainted with the toric derivation,
this is shown exactly as in section 2.2.1: A vev for $Z_{14}$ corresponds to
vevs for $p_6$, $p_7$, $p_9$, a possibility which is consistent with the
symmetry (\ref{invop}). This vevs are forced in the region of FI space
given by $\zeta_1\gg 0$, $\zeta_2=\zeta_3=0$, which is consistent with
(\ref{invozeta}). The variables invariant under the unbroken symmetries
are as in (\ref{varspp}), $x=p_1 p_2^2 p_3 p_8$, $y= p_1 p_3 p_4 p_5^2$,
$z=p_4 p_8$, $w'=p_1 p_2 p_3 p_5 =w/z$, satisfying the relation above.

The final model is an orientifold of type IIB string theory on $xy=zw'^2$
by $\Omega(-1)^{F_L}\rr$, with $\rr$ a geometric $\IZ_2$ transformation
inherited from the action (\ref{orz2z2}) of the orientifold on \z2z2. We
have
\beqa
\rr: \quad x\to -x \;\; ,\;\; y\to y \;\; ,\;\; z\to -z \;\; ,\;\; w'\to w'.
\label{erre1}
\eeqa
This orientifold preserves $\NN=1$ supersymmetry on the D3-branes.

Let us consider the effect of this blow-up in the field theory. The vev
for $Z_{14}$ forces the breaking $SU(n_1)\to SO(n_1)$. The field $Z_{14}$
is swallowed in the process, and the superpotential (\ref{supmodelc})
makes $X_{31}$, $Y_{21}$ and $U^a$ massive. The remaining light fields are
{\small
\beqa
\begin{array}{cccc}
& SO(n_1) & SU(n_2) & \\
X_{13} & \fund & \fund & \\
Y_{12} & \fund & \antifund & \\
Z_{41} & \Yasymm &  & \\
Z_{23} & & \Ysymm & \\
Z_{32}& & \bYasymm & \\
T_a & & \antifund & a=1,\ldots,8
\end{array}
\label{sppone}
\eeqa
}
Integrating out the massive fields, we obtain the superpotential
\beqa
W= \Tr [\, X_{13}^T\, Z_{41}\, Y_{12}\, +\, Y_{12}^T\, X_{13}\, Z_{32}\,
Z_{23}^T\,]\, +\, Z_{23}\, T\, T
\eeqa
This field theory agrees with that arising from the IIA configuration in
figure \ref{fig:orspp2}a. Indeed this is supported by some information from
the T-duality argument. The T-duality between a IIA models with two
NS$'$-branes and one NS-brane to the singularity $xy=zw'^2$ maps the
coordinates 45, 89 to $w'$, $z$. Therefore, the orientifold symmetry imposed
by the O6$'$-planes (reflecting 89, but not 45) T-dualizes to $z\to -z$,
$w'\to w'$, indeed contained in (\ref{erre1}).

\medskip

There is another inequivalent Higgsing that can be performed in the model
C, given by a vev for one of the antisymmetric representations, say
$Z_{41}$. Geometrically, this corresponds again to a blow-up $w'=w/z$,
leading again to an orientifold of $xy=zw'^2$ by $\Omega(-1)^{F_L}\rr$, with
$\rr$ as above (\ref{erre1}). The derivation of this result is different,
though. In the language of section 2.2.1, the blow up induced by the vev for
$Z_{41}$ correspond to vevs for $p_4$, $p_7$, $p_8$ (consistent with
(\ref{invop})). This is achieved in the region of FI terms $\zeta_1\ll 0$,
$\zeta_2=\zeta_3=0$ (consistent with (\ref{invozeta})). The invariant
variables (\ref{vars}) become $x=p_1p_2^2p_3p_9$, $y=p_1p_3p_5^2p_6$,
$z=p_6 p_9$, and $w=p_1p_2p_3p_5p_6p_9$, and the new invariant is
$w'=p_1p_2p_3p_5 =w/z$. They satisfy the relation above.

Given this underlying difference, the field theory on the D3-branes along
this baryonic branch differs from (\ref{sppone}). The group
$SU(n_1)$ breaks to $USp(n_1)$, $Z_{41}$ is swallowed, and $X_{13}$,
$Y_{12}$ get
massive. The remaining light fields are
{\small
\beqa
\begin{array}{cccc}
& USp(n_1) & SU(n_2) & \\
X_{31} & \fund & \antifund & \\
Y_{21} & \fund & \fund & \\
Z_{14} & \Ysymm & & \\
Z_{23} & & \Ysymm & \\
Z_{32}& & \bYasymm & \\
U_a & \fund & & a=1,\ldots,8 \\
T_a & & \antifund & a=1,\ldots,8
\end{array}
\label{spectoomuch}
\eeqa
}
The superpotential is given by
\beqa
W= \Tr [\, X_{31} Z_{14} Y_{21}^T \,+\, Y_{21} X_{31}^T Z_{23} Z_{32}\,]\,
+\, Z_{23}TT \,+\, Z_{14}UU
\label{suptoomuch}
\eeqa
The spectrum suggests the model provides the T-dual of the IIA brane
configuration in figure \ref{fig:orspp2}b. In fact, the orientifold action
on the IIA picture agrees, {\em via} T-duality, with the action on the IIB
side. However the spectrum (\ref{spectoomuch}) has additional states as
compared with (\ref{specfine}), concretely eight chiral flavours of $USp(n_1)$.
The precise IIA configuration to which this model T-dualizes must contain
eight (whole) D6$'$-branes between the NS$'$-branes. Notice how the coupling
between these flavours and the adjoint of $USp(n_1)$ in (\ref{suptoomuch}) is
indeed present in the IIA picture.

There are two important lessons to learn from this example. The first is that
our method has a small caveat. Even though it yields completely consistent
orientifolds of non-orbifold spaces, there is no guarantee that it yields the
{\em simplest} model for a given orientifold action. In the case above, the
IIA picture does not require these D6$'$-branes for consistency, and so
suggests that the IIB D7-branes responsible for the appearance of the flavours
$U$ are not required in the orientifold of the SPP singularity. However, we
know they were actually required for consistency of our starting point, the
model C. The resolution of the puzzle is that in the process of connecting the
orientifold of \z2z2 to the orientifold of the SPP some tadpoles become
non-dangerous, the corresponding RR-flux now being able to escape to infinity
along the infinitely blown-up $\IP_1$. In what follows we will always 
remove this type of additional states whenever they appear.

The second lesson is that in some circumstances one D7-brane in the IIB
side can map to one {\em whole} D6$'$-brane, as in the model above. Notice
that the eight D6$'$-branes in the IIA picture suffer the projection by
the {\em whole} O6$'\,^-$-plane, and so have a $SO(8)$ symmetry, in
agreement with the IIB picture.

Since the two models constructed, (\ref{sppone}) and (\ref{spectoomuch}), are
geometrically the same, but provide different field theories on the probes, we
learn that the orientifold $\Omega (-1)^{F_L}\rr$, with $\rr$ given by
(\ref{erre1}), allows for two different projections ($SO$ and $Sp$) on the
D3-branes.

\medskip

The analysis we have performed can be applied analogously to the remaining
$\IZ_2\times \IZ_2$ orientifolds, models A1, A2 and B. The orientifolds of
$xy=zw^2$ obtained by the blowing-up procedure reproduce all the field
theories proposed in section 3.1, which arise for type IIA configurations
with two NS$'$-branes and one NS-brane. Moreover, the geometric action of the
orientifold projection agrees with the expectations from directly T-dualizing
the orientifold action on the IIA models. The results for the different
orientifolds are shown in the following table.
{\small
\begin{center}
\begin{tabular}{|c||c||c|c|}
\hline
$xyz=w^2$ & Higgsing/Blow-up & $\Omega(-1)^{F_L}\rr$ on $xy=zw'^2$ & Type
IIA configuration \\
\hline\hline
\begin{tabular}{c} A1 \\ $x\to x$, $y\to y$, \\ $z\to z$, $w\to -w$
\end{tabular}
& \begin{tabular}{c} $\langle Z_{14}\rangle_{\smasymm}$ \\
$w'=w/z$  \end{tabular}
& \begin{tabular}{c} $x\to x$, $y\to y$, \\ $z\to z$, $w'\to -w'$
\end{tabular}
& \begin{tabular}{c} Fig.~\ref{fig:orspp1}d \\ (O6$^-$, NS)-NS$'$-O6$^-$
\end{tabular} \\
\hline
\begin{tabular}{c} A2 \\ $x\to x$, $y\to y$, \\ $z\to z$, $w\to -w$
\end{tabular}
& \begin{tabular}{c} $\langle Z_{14}\rangle_{\smsymm}$ \\
$w'=w/z$  \end{tabular}
& \begin{tabular}{c} $x\to x$, $y\to y$, \\ $z\to z$, $w'\to -w'$
\end{tabular}
& \begin{tabular}{c} Fig.~\ref{fig:orspp1}a \\ (O6$^+$, NS)-NS$'$-O6$^+$
\end{tabular} \\
\hline
B & \begin{tabular}{c} $\langle Z_{14}\rangle_{\smasymm}$ \\
$w'=w/z$  \end{tabular}
& \begin{tabular}{c} $x\to -x$, $y\to -y$, \\ $z\to z$, $w'\to -w'$
\end{tabular}
& \begin{tabular}{c} Fig.~\ref{fig:orspp1}b \\ (O6$^+$, NS)-NS$'$-O6$^-$
\end{tabular} \\
\cline{2-4}
\begin{tabular}{c} $x\to -x$, $y\to -y$, \\ $z\to z$, $w\to -w$
\end{tabular}
& \begin{tabular}{c} $\langle Z_{23}\rangle_{\smsymm}$ \\
$w'=w/z$  \end{tabular}
& \begin{tabular}{c} $x\to -x$, $y\to -y$, \\ $z\to z$, $w'\to -w'$
\end{tabular}
& \begin{tabular}{c} Fig.~\ref{fig:orspp1}c \\ (O6$^-$, NS)-NS$'$-O6$^+$
\end{tabular} \\
\hline\hline
C & \begin{tabular}{c} $\langle Z_{14}\rangle_{\smsymm}$ \\
$w'=w/z$  \end{tabular}
& \begin{tabular}{c} $x\to -x$, $y\to y$, \\ $z\to -z$, $w'\to w'$
\end{tabular}
& \begin{tabular}{c} Fig.~\ref{fig:orspp2}a \\
(O6$'$, NS)-NS$'$-O6$'\,^+$ \end{tabular} \\
\cline{2-4}
\begin{tabular}{c} $x\to -x$, $y\to y$, \\ $z\to -z$, $w\to -w$
\end{tabular}
& \begin{tabular}{c} $\langle Z_{41}\rangle_{\smasymm}$ \\
$w'=w/z$  \end{tabular}
& \begin{tabular}{c} $x\to -x$, $y\to y$, \\ $z\to -z$, $w'\to w'$
\end{tabular}
& \begin{tabular}{c} Fig.~\ref{fig:orspp2}b \\
(O6$'$, NS)-NS$'$-O6$'\,^-$ \end{tabular} \\
\hline
\end{tabular}
\end{center}
}

In the last column we indicate the objects involved in the corresponding
IIA configuration (not including their $\IZ_2$ images), ordered as they
appear along the $x^6$ interval. Objects enclosed in parentheses are
located at the same $x^6$ position (for instance, (O6$'$-NS corresponds
to a fork configuration).

\medskip

{\bf Further blow-ups}

In this section we would like to make some comments on further blow-ups. As
explained above, baryonic branches associated to fields $X$ or $Y$ arise from
two blow-ups in the orbifold theory, which become mapped to each other in the
orientifolding process. These can be studied using the toric methods we 
have described, but their net effect is to introduce a variable $y'=y/w^2$ 
(or $x'=x/w^2$). Hence, this type of blow-up leads to orientifolds of smooth
spaces, concretely Taub-NUT spaces. The precise action of the orientifold on
the geometry can be computed as in the examples above. These blow-ups have a
simple interpretation in terms of the T-dual IIA picture,  they correspond to
the removal of the two parallel NS$'$-branes from the configuration.

The resulting field theories are not uninteresting. After the Higgsing,
the IIA configurations in figures \ref{fig:orspp1}b, \ref{fig:orspp1}c 
realize the $\NN=4$ $USp(n)$ and $SO(n)$ gauge theories, respectively, as 
constructed in \cite{uranga}. In the type IIB picture, the transverse 
space to the D3-branes, given by the Taub-NUT $xy'=z$ times a complex 
plane parametrized by $w'$, is modded
out by the orientifold action $\Omega(-1)^{F_L}\rr'$, with $\rr'$ acting as
\beqa
x\to -x \;\; ,\;\; y'\to -y' \;\; ,\;\; z\to z \;\; ,\;\; w'\to -w'
\eeqa
The model in figure \ref{fig:orspp1}a (resp. fig. \ref{fig:orspp1}d) realizes
an $\NN=2$ $SO(n)$ (resp. $USp(n)$) gauge theories with one symmetric (resp.
antisymmetric) hypermultiplet. The $USp(n)$ theory, with four additional
hypermultiplet flavours, has appeared in several papers \cite{probe, uranga}.
In the T-dual IIB picture, the geometric action of the orientifold is
\beqa
x\to x \;\; ,\;\; y'\to y' \;\; ,\;\; z\to z \;\; ,\;\; w'\to -w'
\eeqa
The configurations in figure \ref{fig:orspp2} contain a fork even after the
Higgsing. The models are however non-chiral, and actually exhibit an enhanced
$\NN=2$ supersymmetry. After the Higgsing, the model in figure
\ref{fig:orspp2}b (resp. figure \ref{fig:orspp2}a) realizes a $USp(n)$ 
($SO(n)$) theory with one antisymmetric (symmetric) and four fundamental 
hypermultiplets. In the type IIB picture the corresponding orientifold 
action is
\beqa
x\to -x \;\; ,\;\; y'\to y' \;\; ,\;\; z\to -z \;\; ,\;\; w'\to w'
\eeqa
The $\NN=2$ models above illustrate examples of field theories which can be
realized in two different IIA brane constructions. Our analysis shows that
their type IIB version also correspond to different orientifold projections
\footnote{We thank J.~Erlich and A.~Naqvi for raising this question, 
and for useful discussions on this point.}.

\medskip

A different pattern is obtained if we explore blow-ups of the orientifolds
of $xy=zw'^{\,2}$ associated with vevs for the fields $Z_{ij}$. From the
type IIB perspective, we can use the techniques of section 2.2.1 to show
that the resulting models are always orientifolds of the $\IZ_2$ orbifold.
This can be also understood in the IIA setup, since these blow-ups are
realized as the removal along $x^7$ of the only NS-brane in the configuration,
leaving the two parallel NS$'$-branes untouched.

Thus, our techniques allow to recover (in a somewhat complicated fashion)
some orientifolds of the $\IC^2/\IZ_2$ orbifold, which can actually be directly
constructed, as in \cite{pu} for $\NN=2$ orientifolds, or as in \cite{pru}
for $\NN=1$ orientifolds. This provides a non-trivial consistency check
for our procedure, by comparing the orientifold actions derived from
the blowing-up technique with the orientifold actions used in the direct
construction of the model. We find agreement in all the models considered.

Let us briefly discuss this point in the orientifold of $xy=zw'^2$ by
$\Omega(-1)^{F_L}\rr$, with $\rr$ given in (\ref{erre1}), with the projection
leading to the spectrum (\ref{sppone}). In this case there are two interesting
blow-ups which can be performed. Let us first consider the baryonic branch
corresponding to a vev for $Z_{23}$. It can be seen to correspond to the
blow-up $x'=x/z$, so the resulting space is an orientifold of $x'y=w'^{\,2}$
by $\Omega(-1)^{F_L}$ times the action
\beqa
x'\to x' \;\; ,\;\; y\to y \;\; ,\;\; w'\to w' \;\; ,\;\; z\to -z
\label{agree1}
\eeqa
From the field theory point of view, the vev for $Z_{23}$ yields a final
(non asymptotically free) $\NN=2$ theory with gauge group $SO(n_1)\times
SO(n_2)$ with one bifundamental hypermultiplet. This field theory can be
directly constructed (in analogy with models in \cite{pu}) on the world-volume
of D3-branes at an orientifold obtained by modding out $\IC^3$by ${\IZ_2}
+ \Omega_3 {\IZ_2}$. Here $\Omega_3=\Omega(-1)^{F_L}R_3$ and recall
that $R_3$ flips the sign of $z_3$, and that the generator $\theta$ of
$\IZ_2$ flips the sign of $z_1$, $z_2$. The action of the orientifolding
element $\Omega_3$ on the invariant variables $x=z_1^2$, $y=z_2^2$,
$w=z_1,z_2$, $z=z_3$ (satisfying $xy=w^2$) is exactly (\ref{agree1}), in
agreement with the result from the blowing-up procedure.

A different blow-up would have been achieved if we had given a vev to
$Z_{32}$. This choice can be seen to correspond to the blow-up $y'=y/z$, so
the final orientifold mods $xy'=w'^2$ by $\Omega(-1)^{F_L}$ times the action
\beqa
x\to -x \;\; ,\;\; y\to -y \;\; ,\;\; w'\to w' \;\; ,\;\; z\to -z
\label{agree2}
\eeqa
From the field theory point of view, the vev for $Z_{32}$ give a final
$\NN=2$ theory with group $SO(n_1)\times USp(n_2)$ and one bifundamental
and four fundamental $\fund_2$ hypermultiplets. This theory can be
directly constructed (as in \cite{pu}), by modding out D3-branes in flat
space by the orientifold group ${\IZ_2}+ \Omega_3 \alpha{\IZ_2}$, with
$\alpha:(z_1,z_2)\to(iz_1,-iz_2)$. The action of the orientifold element
$\Omega_3 \alpha\IZ_2$ on the invariant variables is exactly (\ref{agree2}).
These two examples illustrate how the indirect construction of orientifolds
using the blowing-up procedure agrees with the direct construction, when
the latter is available.

\medskip

These two blow-ups are interesting for a different reason. In the T-dual
brane picture (figure \ref{fig:orspp2}a, they correspond to two different
deformations of the fork, corresponding to moving the NS-brane towards
positive $x^7$ (leaving two positively charged O6$'$-planes in the
configuration) or towards negative $x^7$ (leaving two oppositely charged
O6$'$-planes). In the type IIB picture these two blow-ups differ by a flop
transition, as we show below. This was in fact expected, since the distance in
$x^7$ between the NS- and NS$'$-branes in the IIA configuration is mapped to
the K\"ahler size of the $\IP_1$ in the IIB picture.

We can use the techniques in section 2.2.1 to show that the two blow-ups above
differ by a flop transition. The first one corresponds to the region in FI
space $\zeta_1\gg 0$, $\zeta_3+\zeta_2=0$, $\zeta_2\gg 0$. Notice this
choice of FI is consistent with (\ref{invozeta}), so the blow-up indeed
exists in the orientifold model. The D-term equations for the charge
matrix ${\tilde Q}$ (\ref{fincharg}) are solved by $|p_4|^2=\zeta_2$,
$|p_6|^2=\zeta_1$, $|p_7|^2=|p_9|^2=\zeta_1+\zeta_2$ (note that these vevs
are invariant under (\ref{invop})). Using (\ref{xpexplicit}), we see these
vevs imply the blow-up is associated with a vev for $Z_{14}$, $Z_{23}$ in
the field theory. The second blow-up discussed above corresponds to the region
$\zeta_1\gg 0$, $\zeta_3+\zeta_2=0$, $\zeta_2\ll 0$. The D-term equations for
the $p_{\alpha}$ are satisfied by the vevs $|p_6|^2=|p_7|^2=\zeta_1-\zeta_2$,
$|p_8|^2=-\zeta_2$, $|p_9|^2=\zeta_1$, which imply the blow-up corresponds
to vev for $Z_{14}$, $Z_{32}$ in the field theory. As claimed above, the
only difference between both resolutions is the change of sign for
$\zeta_2$, which signals a flop transition, as studied in \cite{flops,muto}.

\section{T-dual models II: Orientifolds of the conifold}

In principle one could try to obtain orientifolds of the conifold by
performing blow-ups in the orientifolds of \z2z2 constructed in section 4.1.
However, after exhaustive exploration whose details we spare here it is
possible to show that this is not possible. This is due to the constraints
(\ref{invozeta}) on the blow-up parameters $\zeta_i$, which forbid precisely
this type of resolution. This has a simple explanation using the relation
between the geometric blow-up and the baryonic Higgsing in the field theory.
Recall that, before orientifolding, the blow-up of \z2z2 to the conifold
corresponds to a vev to exactly two fields arising from different complex
planes (for instance, $Z_{14}$, $Y_{34}$, as in section 2.2). Clearly this
type of vev is not invariant under the orientifold symmetry
(\ref{involution}), which does not relate fields from different complex
planes. Therefore, the associated geometric blow-up mode is `frozen' in the
orientifold model. A particular consequence of this general argument is the
fact, encountered in the previous section, that the orientifolds of the SPP
singularity cannot be resolved to orientifolds of the conifold (instead,
further blow-ups yield orientifolds of the $\IZ_2$ orbifold).

This problem is rather particular to using \z2z2 as starting point. In
this section we show the orientifolds of the conifold are in fact
recovered as partial blow-ups of orientifolds of the \twothree orbifold.

\subsection{The \twothree orbifold and its blow-up to the conifold}

Let us first discuss the model before the orientifold projection. The
action of the $\IZ_2\times \IZ_3$ on $\IC^3$ that we are considering is
actually equivalent to that generated by the order six element $\theta$
\beqa
\theta: (z_1,z_2,z_3)\to (e^{2\pi i\frac{1}{6}} z_1, e^{2\pi i\frac{1}{3}}
z_2, e^{2\pi i\frac{3}{6}})
\label{def}
\eeqa
The orbifold is also knows as $\IZ_6'$. Using the invariant variables
$x=z_1^6$, $y=z_2^3$, $z=z_3^2$, $w=z_1z_2z_3$, the space can also be
described as the hypersurface in $\IC^4$ defined by $xy^2z^3=w^6$.

The gauge group \footnote{In contrast with the $\IZ_2\times \IZ_2$ models,
in this case the issue of the $U(1)$ factors is more subtle, since some of
them have non-zero triangle anomalies. These anomalies are, however,
cancelled by a GS mechanism, as shown in \cite{iru}. A similar comment
applies to the orientifold models below.} on the world-volume of a set of
D3-brane probes in this space is $\prod_{i=1}^6 SU(n_i)\times U(1)$.
There are also eighteen $\NN=1$ matter multiplets, denoted
\beqa
& X_{12}, X_{23}, X_{34}, X_{45}, X_{56}, X_{61} \nonumber\\
& Y_{13}, Y_{24}, Y_{35}, Y_{46}, Y_{51}, Y_{62} \nonumber \\
& Z_{14}, Z_{25}, Z_{36}, Z_{41}, Z_{52}, Z_{63}
\label{specz6p}
\eeqa
The subindices $ij$ indicate the field transforms in the
$(\fund_i,\antifund_j)$ representation. The superpotential is given by
\beqa
W= & \Tr [ X_{12}Y_{24}Z_{41} - X_{12}Z_{25}Y_{51} + X_{23}Y_{35}Z_{52}
- X_{23}Z_{36}Y_{62} + X_{34}Y_{46}Z_{63} - X_{34}Z_{41}Y_{13} + \nonumber\\
& +X_{45}Y_{51}Z_{14} - X_{45}Z_{52}Y_{24} + X_{56}Y_{62}Z_{25}
- X_{56}Z_{63}Y_{35} + X_{61}Y_{13}Z_{36} - X_{61}Z_{14}Y_{46} ] \nonumber
\eeqa
The moduli space of this theory can be constructed explicitly following
the technique in section 2.2.1. Since it does not introduce new conceptual
ingredients, we provide the basic data in the appendix B. Following the
same arguments as in the $\IZ_2\times \IZ_2$ case, they can be used to
reproduce the geometric results mentioned in our arguments below.

\begin{figure}
\centering
\epsfxsize=5.5in
\hspace*{0in}\vspace*{.2in}
\epsffile{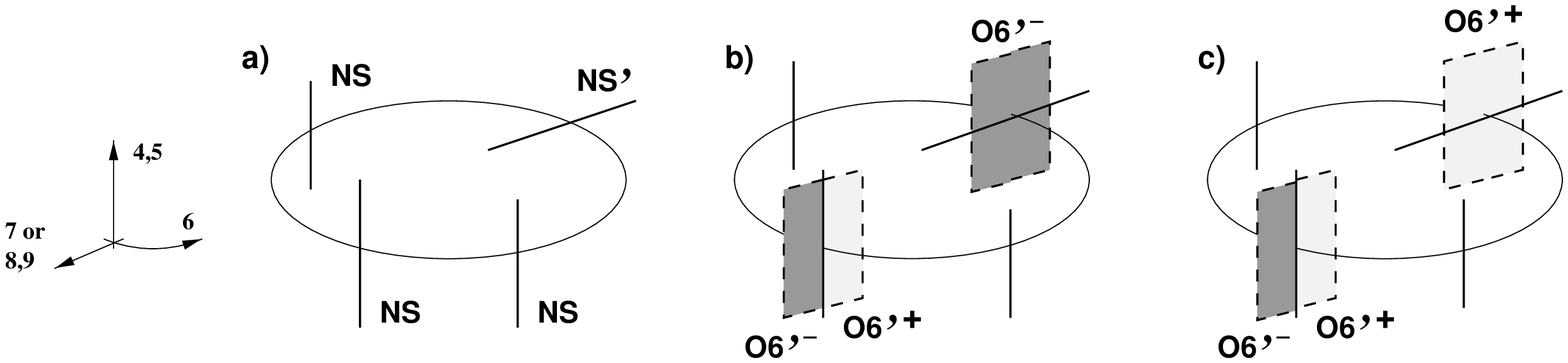}
\caption{\small Type IIA brane configurations related by T-duality to
D3-branes at the $zy=zw^3$ singularity (figure a), and orientifolds
thereof (figures b and c).}
\label{fig:z6prime}
\end{figure}

Since the process of blowing up to the conifold is somewhat involved, we
proceed in two steps. First consider the geometric resolution associated
to a vev for $Z_{63}$, $Z_{41}$ (so we are assuming $n_6=n_3, n_1=n_4$).
From the geometric point of view, it corresponds to resolving the \twothree
orbifold to the singularity $xy=zw^3$, as shown in appendix B. From the
field theory point of view, the gauge group is broken to $SU(n_1)\times
SU(n_2)\times SU(n_3)\times SU(n_5)$, and the fields $Z_{63}$, $Z_{41}$ 
are swallowed. The superpotential couplings give masses to the fields 
$X_{12}$, $X_{34}$, $X_{56}$, $Y_{24}$, $Y_{35}$, and a linear combination 
of $Y_{46}$, $Y_{13}$. The remaining light fields transform as follows
{\small
\beqa
\begin{array}{ccccc}
& SU(n_1) & SU(n_2) & SU(n_3) & SU(n_5) \\
Y & \fund & & \antifund & \\
X_{61} & \antifund & & \fund & \\
Z_{36} & & & {\rm Adj} & \\
X_{23} & & \fund & \antifund & \\
Y_{62} & & \antifund & \fund & \\
Z_{25} & & \fund & & \antifund \\
Z_{52} & & \antifund & &\fund \\
X_{45} & \fund & & & \antifund \\
Y_{51} & \antifund & & & \fund \\
Z_{14} & {\rm Adj.} & & & \\
\end{array}
\eeqa
}
where $Y$ is the linear combination of $Y_{46}$, $Y_{13}$ which remains
massless. The superpotential for these modes is
\beqa
W= \Tr [ Z_{36} X_{61} Y - Z_{36} Y_{62} X_{23} + X_{23} Y_{62} Z_{25} Z_{52}
- Z_{52} Z_{25}  Y_{51} X_{45} + X_{45} Y_{51} Z_{14} - Z_{14} Y X_{61} ]
\nonumber
\eeqa
The fact that this field theory appears on the world-volume of D3-branes
at the singularity $xy=zw^3$ is also supported by the T-duality mentioned
in Section~1.1 \cite{uraconi}. The configuration is mapped to a type IIA
model containing three NS-branes and one NS$'$-brane, with D4-branes
suspended among them. This configuration, which shown in figure
\ref{fig:z6prime}a, allows to easily read off the field theory and
interactions above. The picture is also helpful since it suggest that the
field theory of the conifold is recovered upon giving vevs to {\em e.g.}
$X_{45}$, $X_{23}$, since this corresponds to removing two NS-branes from the
picture, recovering figure~\ref{fig:conspp}. The field theory analysis is
straightforward, and we obtain a gauge group $SU(n_1)\times SU(n_2)$, with
the multiplets $Z_{52}$, $Y$ transforming in the $(\fund,\antifund)$, and the
multiplets $Z_{25}$, $X_{61}$ in the $(\antifund,\fund)$. The superpotential
is
\beqa
W= \Tr[Z_{52}Z_{25}YX_{61}- Z_{52}X_{61}YZ_{25}
]
\eeqa
This field theory agrees with (\ref{speconi}) and (\ref{coniagain}).
It is also a simple matter to find the geometric counterpart of this Higgs
branch, and find that it corresponds to blowing up $xy=zw^3$ to the
conifold singularity (see appendix B).

\medskip

\subsection{Orientifolds of the conifold}

Let us turn to the study of orientifold models. We will consider as our
starting point the orientifold of \twothree obtained by imposing the
orientation reversing projection $\Omega'\equiv \Omega(-1)^{F_L}R_1R_2R_3$
(which preserves $\NN=1$ supersymmetry on the D3-branes. We also choose
the following Chan-Paton matrices
\beqa
&\gamma_{\theta,3}=\diag (e^{\pi i \frac 16} 1_{n_1}, e^{\pi i \frac 36}
1_{n_2}, e^{\pi i \frac 56} 1_{n_3},e^{\pi i\frac 76} 1_{n_4}, e^{\pi
i \frac 96} 1_{n_5}, e^{\pi i \frac{11}{6}}1_{n_6} ) \\
& {\footnotesize \gamma_{\Omega',3}=\pmatrix{
 & & & & & 1_{n_1} \cr
 & & & & 1_{n_2} & \cr
 & & & 1_{n_3}& & \cr
 & & 1_{n_3}& & & \cr
 & 1_{n_2}& & & & \cr
 1_{n_1}& & & & & \cr
} \quad  {\rm or} \quad
\gamma_{\Omega',3}=\pmatrix{
 & & & & & 1_{n_1} \cr
 & & & & 1_{n_2} & \cr
 & & & 1_{n_3}& & \cr
 & & -1_{n_3}& & & \cr
 & -1_{n_2}& & & & \cr
 -1_{n_1}& & & & & \cr}
} \nonumber
\label{infa1}
\eeqa
where the choice of $\gamma_{\Omega',3}$ determines the type of projection
on the D3-branes. This type of models has been studied in \cite{zwart,afiv}
in the compact case. For the symmetric $\gamma_{\Omega',3}$, the resulting
spectrum is
{\small
\beqa
\begin{array}{cccc}
& SU(n_1) & SU(n_2) & SU(n_3) \\
X_{61} & \bYasymm & & \\
X_{12} & \fund & \antifund & \\
X_{23} & & \fund & \antifund \\
X_{34} & & & \Yasymm \\
Y_{51} & \antifund & \antifund & \\
Y_{13} & \fund & & \antifund \\
Y_{24} & & \fund & \fund \\
Z_{52} & & \bYasymm & \\
Z_{25} & & \Yasymm & \\
Z_{14} & \fund & & \antifund \\
Z_{41} & \antifund & & \antifund
\end{array}
\label{specinter}
\eeqa
}
with superpotential
\beqa
W= \Tr [\, X_{12} Y_{24} Z_{41} - X_{12} Z_{25} Y_{51} - X_{23} Y_{24}^T
Z_{52} - X_{23}^T Y_{51} Z_{14} - X_{61} Y_{13} Z_{14}^T - X_{34} Z_{41}
Y_{13} \,]. \nonumber
\eeqa
For the antisymmetric $\gamma_{\Omega',3}$ the spectrum and interactions are
analogous, and can be obtained by replacing the antisymmetric representations
in (\ref{specinter}) by symmetric representations. We will not discuss the
details of this case, and mainly treat the case listed above. Also, since we
are interested in baryonic Higgs branches which exist only for
$n_1=n_2=n_3=n$, from now on we assume equal ranks for all gauge factors.

The orientifold above contains non-vanishing tadpoles which must be cancelled.
Notice that the inconsistency arising from these tadpoles is manifest since
the field theory has non-abelian gauge anomalies. The simplest possibility to
cancel the tadpoles (in the case of equal $n_i$'s)is to add a set of
D7$_2$-branes with Chan-Paton matrices
\beqa
\gamma_{\theta,7_2}=\diag (i1_4,-i1_4)\quad ; \quad
\gamma_{\Omega',7_2}=\pmatrix{ & 1_4 \cr -1_4 & \cr}
\label{mat1}
\eeqa
The symmetry of the matrix $\gamma_{\Omega',7_2}$ is determined by that of
$\gamma_{\Omega',3}$.
States in the 3-7$_2$ sector provide chiral antifundamental multiplets
$T_1^{a}$, $a=1,\ldots, 4$ for $SU(n)_1$, and chiral fundamentals $T_3^a$,
$a=1,\ldots,4$ for $SU(n)_3$. This additional matter precisely cancels the
field theory anomalies. These fields couple to the 33 sector through the
superpotential
\beqa
W= T_1^a Y_{13} T_3^a
\label{supsu4}
\eeqa

For completeness, let us mention that for antisymmetric $\gamma_{\Omega',3}$,
tadpoles (and anomalies) are cancelled by introducing D7$_1$-branes with
Chan-Paton factors
\beqa
\gamma_{\theta,7_1}=\diag (1_4,-1_4)\quad ; \quad
\gamma_{\Omega',7_1}=\pmatrix{ 1_4 & \cr & 1_4 \cr}
\label{mat2}
\eeqa
This gives rise to four chiral fundamental flavours for $SU(n_1)$ and four
anti-fundamental flavours for $SU(n_3)$. They couple to the symmetric
representations $X_{61}$ and $X_{34}$.

In the following we would like to describe the blow-up to the orientifold of
the conifold. In order to interpret geometrically the Higgsing involved, it
will be useful to list here the action of the orientifold symmetry on the
fields (\ref{specz6p}) of the orbifold theory
\beqa
\begin{array}{ccccc}
X_{61} \leftrightarrow X_{61} & \quad & Y_{62} \leftrightarrow Y_{51} &
\quad & Z_{52} \leftrightarrow Z_{52} \\
X_{12} \leftrightarrow X_{56} & \quad & Y_{13} \leftrightarrow Y_{46} &
\quad & Z_{63} \leftrightarrow Z_{41} \\
X_{23} \leftrightarrow X_{45} & \quad & Y_{24} \leftrightarrow Y_{35} &
\quad & Z_{14} \leftrightarrow Z_{36} \\
X_{34} \leftrightarrow X_{34} &\quad & &\quad & Z_{25} \leftrightarrow Z_{25}
\end{array}
\label{invosecond}
\eeqa
The action on the orbifold space, when described as $xy^2z^3=w^6$, is
\beqa
x\to x \;\; ,\;\; y\to -y \;\; ,\;\; z\to z \;\; ,\;\; w\to -w
\label{actz6p}
\eeqa
Finally, it is important for the geometric computations to recall the
constraints imposed on the blow-up parameters $\zeta_i$, which appear as
FI terms or vevs for baryonic operators in the field theory
\beqa
\zeta_1=-\zeta_6 \;\; ,\;\; \zeta_2=-\zeta_5 \;\; ,\;\; \zeta_3=-\zeta_4.
\eeqa

\medskip

Let us turn to describing the blowing-up process. Consider a vev for the
field $Z_{41}$. Its geometric interpretation is straightforward by regarding
it as giving identical vevs to the fields $Z_{41}$, $Z_{63}$ in the orbifold
theory, a process which we studied above. This shows that this blow-up will
resolve the orientifold of \twothree to an orientifold of $xy=zw^3$. Following
the action (\ref{actz6p}) through the blowing-up up process, the final
orientifold acts on $xy=zw^3$ as
\beqa
x\to x \;\; ,\;\; y\to -y \;\; ,\;\; z\to z \;\; ,\;\; w\to -w
\eeqa
From the field theory point of view, the effect of the Higgsing triggered
by this resolution is to break the gauge group to $SU(n)_1\times
SU(n)_2$. The fields $X_{12}$, $Y_{24}$, $X_{34}$ and the antisymmetric
part of $Y_{13}$ become massive. The remaining light fields are
{\small
\beqa
\begin{array}{cccc}
& SU(n)_1 & SU(n)_2 & \\
X_{23} & \fund & \fund & \\
Y_{51} & \antifund & \antifund & \\
X_{61} & \bYasymm & & \\
Y_{13} & \Ysymm & & \\
T^a & \antifund & & a=1,\ldots, 8 \\
Z_{52} & & \bYasymm & \\
Z_{25} & & \Yasymm & \\
Z_{14} & {\rm Adj} &
\end{array}
\label{sporz6p}
\eeqa
}
where the eight chiral flavours $T^a$ arise from the former $T_1$ and
$T_3$. The superpotential is
\beqa
W=\Tr [\,Y_{51} X_{23}^T Z_{52} Z_{25} + X_{23}^T Y_{51} Z_{14} + X_{61}
Y_{13} Z_{14} + Y_{13} T^a T^a \,]
\eeqa
This field theory is realized by a type IIA configuration shown in figure
\ref{fig:z6prime}b. It can be obtained by introducing O6$'$-planes in figure
\ref{fig:z6prime}a. Using the antisymmetric version of $\gamma_{\Omega',3}$,
one can recover the IIA configuration in figure~\ref{fig:z6prime}c in
complete analogy.

Before proceeding any further, we would like to comment on an interesting
point, concerning the origin of the eight fundamental flavours in the
final field theory. Recall they arise from the eight half D6$'$-branes in
the IIA side. On the type IIB model corresponding to symmetric
$\gamma_{\Omega',3}$, however, these flavour arise from just {\em four}
D7$_2$ branes. The matrices (\ref{mat1}) define four D7$_2$-branes and
their mirror images, leading to a $SU(4)$ gauge symmetry (global symmetry
of the superpotential (\ref{supsu4})). This seems not to agree with 
previous statements identifying one whole IIB D7-brane with one half IIA 
D6$'$-brane \cite{kg,pru}. Also, this structure of IIB D7-branes is rather 
different with the one we had proposed in Section~4 as the T-dual of the 
fork configuration. However, a careful analysis of the geometry of the blow-up
shows there is no contradiction. Concretely, the D7$_2$-branes in the initial
model span the direction $x$ in $xy^2z^3=w^6$. As shown in the appendix B,
the blowing-up process implies the introduction of a new variable $x'$
related to $x$ by $x=x'^2z$, which is a one-to-two map. Consequently each of
the original D7-branes wraps {\em twice} the coordinate $x'$, and we end
up with eight D7-branes spanning the direction $x'$ of the space $x'y=zw'^3$.
This description is readily seen to agree with similar geometric realizations
of the fork configuration in Section~4. This interpretation is also  
supported by the fact that for the choice of antisymmetric 
$\gamma_{\Omega',3}$, the initial model already contains eight D7$_1$-branes. 
However, since they do not span the direction $x$, they do not suffer this 
`doubling' process, and provide the right number of T-dual half D6$'$-branes.  
A related interesting feature of this blowing-up process is the accidental
enhancement of the D7-brane gauge symmetry from $SU(4)$ (or $SO(4)^2$
in the case of antisymmetric $\gamma_{\Omega',3}$) to $SO(8)$. It would be
nice to gain some additional insight into this issue.

\medskip

The final blow-up to the orientifold of the conifold corresponds to giving
a vev to $X_{23}$. Notice that it has a natural interpretation in the IIA
picture, as the removal of two ($\IZ_2$ related) NS-branes, to recover
figure \ref{fig:orconi1}b. In our geometric description, the effect of
such vev is equivalent to the effect of an identical vev for the
fields $X_{23}$ and $X_{45}$ in the orbifold model (notice they are
related by (\ref{invosecond}). Our analysis above implies that this
resolves the orientifold of $xy=zw^3$ to an orientifold of $xy=zw$. The
precise geometric action on this last space is given by
\beqa
x\to x \;\; ,\;\; y\to -y \;\; ,\;\; z\to z \;\;, \;\; w\to -w
\label{actorcon}
\eeqa
From the field theory point of view, the Higgsing implied by the vev above
breaks the gauge group to a single $SU(n)$ factor, with  the following
matter content
\beqa
& X_{61} : \;\; \bYasymm \quad ; \quad
& Y_{13} : \;\; \Ysymm \quad ; \quad
T^a : \;\; \antifund \quad a=1,\ldots,8 \nonumber \\
& Z_{52} : \;\; \bYasymm \quad ; \quad
& Z_{25} : \;\; \Yasymm
\eeqa
with superpotential
\beqa
W=\Tr [\, - X_{61}Y_{13}Z_{52}Z_{25} \,] + Y_{13} T T
\eeqa

This indeed reproduces the field theory (\ref{sporcon}), arising from the
IIA brane configuration in figure \ref{fig:orconi1}b. The alternative
projection on D3-branes reproduces the field theory realized by the
configuration in figure \ref{fig:orconi1}a. Finally, let us mention that
the geometric action  (\ref{actorcon}) is also in agreement with
expectations from directly T-dualizing the orientifold action on the IIA
side.

\medskip

{\bf Alternative blow-ups}

Clearly, many other blow-ups of the initial orientifold of \twothree are
possible, and performing an exhaustive exploration is outside the point of
this paper. Suffice it to say that in all cases we have found results
consistent with our proposals in previous sections. For instance, the
orientifold of $xy=zw^3$ constructed above (with spectrum (\ref{sporz6p}))
can be blown up an orientifold of the SPP. More concretely, there are two
inequivalent ways to perform such a blow-up, differing by a flop
transition. The resulting models correspond to introducing a new $\IP_1$
parametrized by either $y'=y/w$ or $x'=x/w$. The two resulting models
correspond to modding out the space $xy=zw^2$ by $\Omega (-1)^{F_L}$ times
the geometric actions
\beqa
x\to \pm x \;\; ,\;\; y\to\pm y \;\; ,\;\; z\to z \;\; ,\;\; w\to -w
\label{actfin}
\eeqa
These blow-ups correspond to vevs for $X_{61}$ (for upper signs) or
$Y_{13}$ (for lower signs).
It is straightforward to obtain the field theories after the Higgsing. In
fact, notice that this Higgsings correspond to removing the NS-brane stuck
at the fork in figure  \ref{fig:z6prime}b. The resulting brane configurations
are, respectively, given in figures \ref{fig:orspp1}d and \ref{fig:orspp1}c
(up to an irrelevant relabeling of primed and unprimed objects), so the
spectrum of the final theories can be read off from them. The geometric
action (\ref{actfin}) is in perfect agreement with the T-duality with
these IIA configurations, as can be checked using the actions we had
proposed in section 4.2.

A different possibility is to blow-up the orientifold of \twothree to
an $\NN=1$ orientifold of $\IC^3/\IZ_3$. The direct construction of the
latter in \cite{pru} provides a new non-trivial check of our 
procedure.
Again, we find agreement between the orientifolds arising after the
blow-up, the spectra resulting after the corresponding Higgsing, and the
T-dual IIA brane configurations.

\section{Final comments}

Here we would like to make some comments on the more general case of type
IIA brane configurations with $k$ NS-branes, $k'$ NS$'$-branes, D4-branes
and O6$'$-planes. The possible models can be easily classified in several
families, depending on the parity of the numbers $k$, $k'$, and the
location of the NS fivebranes with respect to the O6$'$-planes.
Within each family, there are different field theories depending on the
ordering of the NS- and NS$'$-branes in the direction 6.

The type IIB T-dual configurations correspond to a set of D3-branes
sitting at an orientifold of $xy=z^kw^{k'}$, with orientifold action
$\Omega(-1)^{F_L}$ times a geometric $\IZ_2$ symmetry. It is a simple
matter to propose suitable  $\IZ_2$ geometric actions corresponding to the
different type IIA models. The results are shown in table \ref{table1}. The
geometric action of the IIB orientifold on $z$, $w$ follows easily by
T-dualizing the action of the IIA orientifold on 89, 45 respectively. The
action on $x$, $y$ can be determined from other requirements (the action
must be a symmetry of the singularity, the holomorphic 3-form must be odd
under it), and consistency with known results (the orientifolds of the
conifold and suspended pinch point studied in this paper, as well as the
limiting cases where the singularity is actually an orbifold of flat
space. Thus, when $k=0$ the T-duals proposed in the table agree with those
found in \cite{pu}, while for $k'=0$ the T-duals agree with those proposed
in \cite{pru}).

{\footnotesize
\begin{table}[t!]
\renewcommand{\arraystretch}{1.25}
\begin{center}
\begin{tabular}{|c||c||c||c|}
\hline
 & IIB orientifold of $xy=z^k w^{k'}$ & IIA brane configuration  \\
\hline\hline
$k$ odd & $x\to x$, $y\to -y$ $^*$
& (O6$'^{\,+}$, NS$'$)-\nsp-$\ldots$-\nsp-(NS, O6$'$) \\
\cline{3-3}
$k'$ odd & $z\to -z$, $w\to w$
& (O6$'^{\,-}$, NS$'$)-\nsp-$\ldots$-\nsp-(NS, O6$'$) \\
\hline\hline
         & $x\to x$, $y\to y$ $^*$
& O6$'^{\,+}$-\nsp-$\ldots$-\nsp-(NS$'$, O6$'^{\,+}$) \\
\cline{3-3}
$k$ even & $z\to -z$, $w\to w$
& O6$'^{\,-}$-\nsp-$\ldots$-\nsp-(NS$'$, O6$'^{\,-}$) \\
\cline{2-3}
$k'$ odd & $x\to -x$, $y\to -y$ $^*$
& O6$'^{\,+}$-\nsp-$\ldots$-\nsp-(NS$'$, O6$'^{\,-}$) \\
\cline{3-3}
         & $z\to -z$, $w\to w$
& O6$'^{\,-}$-\nsp-$\ldots$-\nsp-(NS$'$, O6$'^{\,+}$) \\
\hline\hline
$k$ odd   & $x\to x$, $y\to -y$ $^*$
& O6$'^{\,+}$-\nsp-$\ldots$-\nsp-(NS, O6$'$) \\
\cline{3-3}
$k'$ even & $z\to -z$, $w\to w$
& O6$'^{\,-}$-\nsp-$\ldots$-\nsp-(NS, O6$'$) \\
\hline\hline
& $x\to x$, $y\to y$ $^*$
& O6$'^{\,+}$-\nsp-$\ldots$-\nsp-O6$'^{\,+}$ \\
\cline{3-3}
& $z\to -z$, $w\to w$
& O6$'^{\,-}$-\nsp-$\ldots$-\nsp-O6$'^{\,-}$ \\
\cline{2-3}
& $x\to -x$, $y\to -y$
& O6$'^{\,+}$-\nsp-$\ldots$-\nsp-O6$'^{\,-}$ \\
& $z\to -z$, $w\to w$  & \\
\cline{2-3}
& $x\to x$, $y\to y$ $^*$   &
(O6$'^{\,+}$, 
NS$'$)-\nsp-$\ldots$-\nsp-(NS$'$, O6$'^{\,+}$) \\
\cline{3-3}
$k$ even
& $z\to -z$, $w\to w$  &
(O6$'^{\,-}$, NS$'$)-\nsp-$\ldots$-\nsp-(NS$'$, O6$'^{\,-}$) \\
\cline{2-3}
$k'$ even
& $x\to -x$, $y\to -y$ &
(O6$'^{\,+}$, NS$'$)-\nsp-$\ldots$-\nsp-(NS$'$, O6$'^{\,-}$) \\
& $z\to -z$, $w\to w$  & \\
\cline{2-3}
& $x\to x$, $y\to y$   & (O6$'$, NS)-\nsp-$\ldots$-\nsp-(NS, O6$'$) \\
& $z\to -z$, $w\to w$  & (parallel forks) \\
\cline{2-3}
& $x\to -x$, $y\to -y$ & (O6$'$, NS)-\nsp-$\ldots$-\nsp-(NS, O6$'$) \\
& $z\to -z$, $w\to w$  & (antiparallel forks) \\
\hline
\end{tabular}
\end{center}
\caption{\small The first column of the table shows the type IIB 
orientifolds
T-dual to the type IIA brane configurations shown in the second. The
objects in the IIA side are listed as ordered in the coordinates $x^6$.
Objects enclosed in parentheses are located at the same $x^6$ position.
\nsp denotes a NS- or a NS$'$-brane. Notice that we have not
allowed for configurations with coincident fivebranes, since their field
theory interpretation is unclear.}
\label{table1}
\end{table}
}

The geometric actions in table~\ref{table1}, however, do not completely
define the IIB orientifold. In fact, we see that different type IIA
configurations may correspond to the same geometric action. As we know
from our experience, some of the orientifolds allow for two possible
projections ($SO$ and $Sp$) on the D3-branes Chan-Paton factors. This is
not {\em a priori} obvious from the type IIB side unless one has a more
explicit definition of the orientifold. Our procedure of blowing up
orientifolds of orbifolds to obtain orientifolds of non-orbifold
singularities precisely amounts to such a definition, and this allowed us
to claim the existence of these two projections in some models of Section 4.2.
In the general case, T-duality predicts the existence of such choice in
the orientifolds marked with an asterisk in table \ref{table1}. This
choice should distinguish the T-dual IIB orientifolds of the two IIA
configurations shown in the second column.

Another subtle point concerning the precise definition of the IIB orientifold
is the action on the closed string modes localized at the singularity (the
analogs of the twisted modes in the orbifold case). In particular, some
non-orbifold singularities seem to have the analog of $\IZ_2$ twisted sectors,
in that the orientifold maps these modes to themselves. In the orientifolds of
orbifold singularities it is known that there are two possible projections for
these sectors \cite{polchinski}. The orientifold of non-orbifold spaces at
hand show a similar feature. This choice should distinguish the IIB
orientifolds with identical geometric action, but whose claimed T-dual IIA
configurations differ by the location of fivebranes with respect to the
O6$'$ planes.

Let us mention some further evidence supporting the existence of the mentioned
choices in the definition of these orientifolds. In order to do that, notice
that any IIA configuration in table~\ref{table1} without NS$'$-branes stuck at
the O6$'$-planes can be deformed by rotating the NS$'$-branes in 45-89 in a
way consistent with the orientifold projection. Such models can be thus
rotated to models with only NS-branes, D4-branes and O6$'$ branes, and no
NS$'$-branes. The T-duals of these models are D3-branes at $\NN=1$
orientifolds of $\IC^2/\IZ_{k+k'}$ orbifolds \cite{pru}. In fact, the 
whole rotation
process can be followed in the type IIB T-dual picture: for generic rotation
angles, the T-dual is given by an orientifold of $xy = z^k\prod_{i=1}^{k'/2}
(z-\alpha_i w)(z+\alpha_i w)$, which interpolates between the space
$xy=z^kw^{k'}$ (for $\alpha_i=\infty$) and the orbifold $xy=z^{k+k'}$ (for
$\alpha_i=0$). These orientifolds of $\IC^2/\IZ_N$ orbifolds have been
constructed in \cite{pru}, where it was shown that the different choices in
the definition of the IIB orientifold action reproduce the different
possible IIA configurations. Table~\ref{table1} suggests that this feature
is preserved along the deformation from the orbifold space to
$xy=z^kw^{k'}$.

Analogously, models without NS-branes stuck at O6$'$-planes can be rotated
to models without NS-branes. The T-duals of these models are the $\NN=2$
orientifolds of $\IC^2/\IZ_N$ constructed in \cite{pu}. The only models
which cannot be rotated to more familiar configurations are those with
stuck NS- and NS$'$-branes, {\em i.e.} the case of odd $k$, $k'$.

There is a last comment we would like to make concerning this process of
brane rotation. Given that it allows to relate orientifolds of these
non-orbifold spaces to well-known orientifolds of orbifold spaces, it
might be possible to use it to provide a precise definition, and a
computational tool to analyze the former. This would allow the analysis of
large families of such IIB orientifolds with little effort. In fact this
is the situation suggested by the T-duality with IIA configurations: many
properties of the rotated model are related to properties of the
unrotated one. However, from the IIB perspective there are several
difficulties in carrying out this proposal. For instance, it is not clear
how the effect of the deformation of the orbifold appears in the field
theory of the D3-brane probes. The naive proposal is that some adjoint
matter becomes massive and should be integrating out, leaving an effective
theory corresponding to the field theory of D3-branes in the deformed
space. However reasonable this may seem, in most cases this procedure does
not give the correct answer \footnote{This is true even without orientifold
projections, the simplest example being the deformation of the
$\IC^2/\IZ_3$ orbifold to $xy=zw^2$. The procedure of integrating out the
massive adjoints gives the right matter content, but an incorrect
superpotential. This fact has been known to I.~Klebanov and E.~L\'opez for
some time, and we are grateful to them for discussion on this point.}.
Thus, until a better understanding of the field theory manifestation of
such deformations is achieved, the rotation procedure cannot be claimed to
be a good tool to analyze the non-orbifold spaces.

\medskip

We hope the examples we have worked out illustrate the basic features of
the T-duality for IIA configurations with NS- and NS$'$-branes. We also
expect these results to find applications in other contexts, like for
instance the construction of orientifolds of compact Calabi-Yau varieties.
Clearly many directions remain to be explored.

\begin{center}
{\bf Acknowledgements}
\end{center}

It is our pleasure to thank J.~Erlich, A.~Hanany, L.~E.~Ib\'a\~nez,
B.~Janssen,A.~Karch, P.~Meessen and A.~Naqvi for useful conversations.
A.~M.~U. is grateful to  G.~Aldazabal and D.~Badagnani for their insights 
into orientifold constructions, and also to M.~Gonz\'alez for kind 
encouragement and support, and to the Center for Theoretical Physics at 
M.~I.~T. for hospitality. The research of J.~P. is supported by the US. 
Department of Energy under Grant No. DE-FG02-90-ER40542. The research of 
R.~R. is supported by the Ministerio de Educaci\'on y Cultura (Spain) 
under a FPU Grant. The research of A.~M.~U. is supported by the Ram\'on 
Areces Foundation (Spain).

\newpage

\appendix

\section{Appendix: Tadpole calculation of $Z_2 \times Z_2$ orientifold}

Consider the orientifold projection
\begin{equation}
(1+\theta+\omega+\theta\omega)(1+\alpha\Omega')
\end{equation}
with the convention of (\ref{eq:oc}). This is the orientifold action of
Models B and C. In order to obtain the convention of the Model B adopted
in the main text, we have to replace $\alpha$ by $\beta$ and interchange
$\theta$ with $\theta\omega$ in the following formulae. We can calculate
the tadpole starting from the compact $T^6/(\IZ_2 \times \IZ_2)$ orientifold
and then taking the non-compact limit. In this limit, we can ignore the
untwisted tadpole and the twisted sector tadpole inversely proportional to
one of the volume factors of the three tori. Typically we acquire
$\frac{1}{4\sin^2(2\pi b_i)}$ factor in the non-compact limit if an
orientifold action is non-trivial on the $i^{th}$ torus, $z_i \rightarrow
e^{2\pi ib_i} z_i$. This comes from the momentum modes along the $i^{th}$
torus, which become continuous in the infinite volume limit\cite{gjdp}.
 One subtlety of
$\IZ_2 \times \IZ_2$ is that this factor can diverge if corresponding
orientifold action acts trivially on the $i^{th}$ plane. This means that the
twisted tadpole of our interest is proportional to the volume $V_i$ of the
$i^{th}$ torus, diverging in the non-compact limit. Such tadpoles should
vanish in a consistent configuration. In general, one must consider only
the twisted tadpoles proportional to $V_i$ and tadpoles independent of any 
$V_i$. Hence, in the $\IZ_2\times \IZ_2$ case, we should consider the 
tadpoles proportional to $V_1$, $V_2$ and $V_3$, and the tadpoles of order 
1 with no volume dependence. We turn to each case in the following. 

{\bf i)} Tadpole proportional to $V_2$

This comes from the subset of the orientifold action
\begin{equation}
(1+\theta)(1+\alpha\omega\Omega')
\end{equation}
if we ignore the untwisted tadpole from the cylinder amplitude of
D7$_1$, D7$_3$-branes.  
However we should keep all the Klein-bottle and M\"obius
amplitude since all of them have the same volume dependence $V_2$. 
The amplitude is calculated in the usual manner following\cite{gjdp} and 
the result is
\begin{eqnarray}
KB: & & 64\pm 64 \,\,\,\,\, (\alpha\omega\Omega') \\
    & & 64\pm 64 \,\,\,\,\, (\alpha\theta\omega\Omega') \nonumber \\
M:  & & \sum_{i=1,3} -16 \Tr (\gamma^{-1}_{\alpha\omega\Omega', 7_i}
\gamma^{T}_{\alpha\omega\Omega', 7_i})
-16 \Tr (\gamma^{-1}_{\alpha\omega\theta\Omega', 7_i}
\gamma^{T}_{\alpha\omega\theta\Omega', 7_i}) \nonumber \\
C: & & (\Tr \gamma_{\theta, 7_1}-\Tr \gamma_{\theta, 7_3})^2. \nonumber
\end{eqnarray}
On each column of the Klein bottle amplitude, the first 64 comes from
the untwisted sector and the second 64 comes from the $\theta$-twisted
sector. For Model C, we choose + sign for the twisted sector contribution. 
In order to have factorization, one must require  
\begin{equation}
\Tr (\gamma^{-1}_{\alpha\omega\theta\Omega', 7_i}
\gamma^{T}_{\alpha\omega\theta\Omega', 7_i})
=\Tr (\gamma^{-1}_{\alpha\omega\Omega', 7_i}
\gamma^{T}_{\alpha\omega\Omega', 7_i})=\pm\Tr\gamma_{\theta, 7_i}
\end{equation}
with $+$ sign for D7$_1$ branes and $-$ sign for D7$_3$. 
Thus we have
\begin{equation}
-\Tr\gamma_{\theta, 7_1}+\Tr\gamma_{\theta, 7_3}+16=0.
\end{equation}
For Model B, we choose the $-$ sign for the twisted sector 
contribution in
the Klein bottle amplitude and use
\begin{equation}
\Tr (\gamma^{-1}_{\alpha\omega\theta\Omega', 7_i}
\gamma^{T}_{\alpha\omega\theta\Omega', 7_i})
=-\Tr (\gamma^{-1}_{\alpha\omega\Omega', 7_i}
\gamma^{T}_{\alpha\omega\Omega', 7_i})
\end{equation}
with $i=1,3$ to make M\"obius amplitude vanishing. The only remaining 
piece is the cylinder amplitude, whose cancellation imposes
\begin{equation}
-\Tr\gamma_{\theta, 7_1}+\Tr\gamma_{\theta, 7_3}=0.
\end{equation}

Now we turn to the other tadpoles. The following analysis holds true both
for Model B and Model C.

{\bf ii)} Tadpoles proportional to $V_1$

Only cylinder amplitude contributes and we have
\begin{equation}
\Tr \gamma_{\omega, 7_2}-\Tr \gamma_{\omega, 7_3}=0.
\end{equation}

{\bf iii)} Tadpoles proportional to $V_3$

Only cylinder amplitude of D7$_1$, D7$_2$ contributes. We have
\begin{equation}
\Tr \gamma_{\theta\omega, 7_1}-\Tr \gamma_{\theta\omega, 7_2}=0.
\end{equation}

{\bf iv)} Tadpoles of order 1

Only the Klein bottle and M\"obius amplitudes contribute.
For Klein bottle, we evaluate the $\omega$ and $\theta\omega$ twisted 
amplitudes with insertions of $\alpha\Omega'$, $\alpha\theta\Omega'$,
$\alpha\omega\Omega'$, $\alpha\theta\omega\Omega'$. If the action of a 
twist acting along with $\Omega$ is denoted $z_i\to e^{2\pi i b_i} z_i$, 
the  amplitude for the $\omega$ twisted sector is proportional to $\sin 
2\pi b_2 \sin 2\pi b_3$. 
Since $\sin 2\pi b_2=0$ for all the orientifold action to be considered
in \z2z2, Klein-bottle amplitude vanishes for $\omega$ twisted sector.
For $\theta\omega$ twisted sector the amplitude is proportional to
$\sin 2\pi b_1 \sin 2\pi b_2$, and vanishes for the same reason.

For the M\"obius amplitude, we evaluate $\alpha\Omega', 
\alpha\theta\Omega'$ for
D7$_1$, D7$_3$ branes and $\alpha\omega\Omega', \alpha\theta\omega\Omega'$ 
for D7$_2$ branes. If we T-dualize along $z_1$, $z_2$ plane for D7$_3$, along 
$z_2$, $z_3$ plane for D7$_1$ and T-dualize along $z_1$, $z_3$ planes for 
D7$_2$, we end up with $\alpha\theta\omega\Omega'$, $\alpha\omega\Omega'$ 
for 
D3-branes. Denoting by $\theta'$ the twist $z_i\to e^{2\pi i v_i} z_i$ 
that 
accompanies $\Omega$ in these amplitudes, the M\"obius amplitude 
$\theta'\Omega$ for $D3$ brane is proportional to
\begin{equation}
\prod sign(2\pi v_i) \cos \pi v_i
\end{equation}
In order to obtain the answer for D7$_1$, D7$_3$, we put back the zero 
mode
(continuos momemtum mode) contribution in the $z_1$, $z_2$ planes and 
$z_2$, $z_3$ planes, respectively. The amplitude is proportional to
\begin{equation}
\frac{-\cos^2 \frac{\pi}{4}}{\sin^2\frac{3\pi}{4}\sin^2\frac{\pi}{2}}
\Tr (\gamma^{-1}_{\alpha\Omega', 7_i}\gamma^{T}_{\alpha\Omega', 7_i})
+\frac{\cos^2 \frac{\pi}{4}}{\sin^2\frac{3\pi}{4}\sin^2\frac{\pi}{2}}
\Tr (\gamma^{-1}_{\alpha\theta\Omega', 7_i}
\gamma^{T}_{\alpha\theta\Omega', 7_i})
\end{equation}
with $i=1, 3$. 
The amplitude vanishes if
\begin{equation}
\Tr (\gamma^{-1}_{\alpha\Omega', 7_i}\gamma^{T}_{\alpha\Omega', 7_i})
=\Tr (\gamma^{-1}_{\alpha\theta\Omega', 7_i}
\gamma^{T}_{\alpha\theta\Omega', 7_i}) \,\,\,\, i=1,3.
\end{equation}
Similarly,  we obtain for D7$_2$ branes
\begin{equation}
\Tr (\gamma^{-1}_{\alpha\omega\Omega', 7_2}
\gamma^{T}_{\alpha\omega\Omega', 7_2})
=\Tr (\gamma^{-1}_{\alpha\theta\omega\Omega', 7_2}
\gamma^{T}_{\alpha\theta\omega\Omega', 7_2}).
\end{equation}
The tadpole cancellation conditions for models B and C are collected 
in expression (\ref{fintad}) for convenience of the reader.

\medskip

For Model A, the only twisted tadpoles generated by the Klein bottle 
amplitude are inversely proportional to one of the volume factor $V_i$ or, 
proportional to $V_i$ but with vanishing coefficient. This comes from the 
fermionic zero mode factor $\sin 2\pi b_1 \sin 2\pi b_2$ for 
$\theta\omega$ twisted sector and similar factors for the others. The
M\"obius amplitude also vanishes, as we presently explain. If we consider 
D7$_3$ branes, we should evaluate $\omega\Omega'$, $\theta\Omega'$, 
$\Omega'$. This is proportional to 
\begin{equation}
(\prod sign(\sin2\pi v_i))\sin \pi v_1 \sin \pi v_2 \cos \pi v_3
\end{equation}
where $v_i$ defines the twist acting along with $\Omega$. 
Using the twists appearing in the model, these amplitudes all vanish. 
Similar reason holds for D7$_1$- and D7$_2$-branes.

Finally, the cylinder amplitudes give the condition 
\begin{eqnarray}
\Tr \gamma_{\theta\omega, 7_1} -\Tr \gamma_{\theta\omega, 7_2}&=& 0  \\
\Tr \gamma_{\omega, 7_2} -\Tr \gamma_{\omega, 7_3}&=& 0 \nonumber \\
\Tr \gamma_{\theta, 7_3} -\Tr \gamma_{\theta, 7_1}&=& 0 \nonumber 
\end{eqnarray}

To summarize, for Model A,B, and C we obtain the following conditions
\begin{eqnarray}
\Tr \gamma_{\theta\omega, 7_1} -\Tr \gamma_{\theta\omega, 7_2}&=& 0  
\,\,\,\,\,\, (ABC)\\
\Tr \gamma_{\omega, 7_2} -\Tr \gamma_{\omega, 7_3}&=& 0 \,\,\,\,\, 
(ABC) \nonumber \\
\Tr \gamma_{\theta, 7_3} -\Tr \gamma_{\theta, 7_1}&=& 0 \,\,\,\,\, 
(AB) \nonumber \\  
\Tr \gamma_{\theta, 7_3} -\Tr \gamma_{\theta, 7_1}
&=& 16 \,\,\,\,\, (C) \nonumber 
\end{eqnarray}
\begin{eqnarray}
\Tr (\gamma^{-1}_{\alpha\Omega', 7_i}\gamma^{T}_{\alpha\Omega', 7_i})
&=&\Tr (\gamma^{-1}_{\alpha\theta\Omega', 7_i}
\gamma^{T}_{\alpha\theta\Omega', 7_i})  \,\,\,\,\, (BC)\\
\Tr (\gamma^{-1}_{\alpha\omega\Omega', 7_2}
\gamma^{T}_{\alpha\omega\Omega', 7_2})
&=&\Tr (\gamma^{-1}_{\alpha\theta\omega\Omega', 7_2}
\gamma^{T}_{\alpha\theta\omega\Omega', 7_2})  
\,\,\,\,\, (BC)  \nonumber \\
\Tr (\gamma^{-1}_{\alpha\omega\theta\Omega', 7_i}
\gamma^{T}_{\alpha\omega\theta\Omega', 7_i})
&=&-\Tr (\gamma^{-1}_{\alpha\omega\Omega', 7_i}
\gamma^{T}_{\alpha\omega\Omega', 7_i})  \,\,\,\,\, (B) \nonumber \\
\Tr (\gamma^{-1}_{\alpha\omega\theta\Omega', 7_i}
\gamma^{T}_{\alpha\omega\theta\Omega', 7_i})
&=&-\Tr (\gamma^{-1}_{\alpha\omega\Omega', 7_i}
\gamma^{T}_{\alpha\omega\Omega', 7_i})=\pm \Tr \gamma_{\theta, 7_i} 
 \,\,\,\,\, (C) \nonumber 
\label{fintad}
\end{eqnarray}
where $i=1,3$ and $+$ sign for D7$_1$ branes and $-$ sign for D7$_3$ 
on the last line. 
For Model A and B, the above conditions can be satisfied without 
introducing any D7 branes while for Model C we can introduce only 
D7$_3$ branes to satisfy the constraints using the Chan-Paton matrices 
appearing (\ref{eq:occ}). Certainly there are various other solutions. 
However, they are different from this minimal solution by non-chiral 
matter contents. 

\section{Appendix: Moduli space of \twothree}

Following the procedure explained in section 2.2.1, the initial fields
(\ref{specz6p}) can be parametrized in terms of seventeen fields
$p_{\alpha}$, as follows
\beqa
\begin{array}{lll}
X_{12}= p_1 p_3 p_4 p_5 p_8 p_9 &
Y_{13}= p_4 p_5 p_8 p_{11} p_{15} p_{16} &
Z_{14}= p_5 p_9 p_{14} p_{16} p_{17}
\\
X_{23}= p_3 p_6 p_8 p_{10} p_{15} p_{16} &
Y_{24}= p_7 p_{10} p_{11} p_{15} p_{16} p_{17} &
Z_{25}= p_6 p_{13} p_{14} p_{16} p_{17}
\\
X_{34}= p_1 p_3 p_7 p_9 p_{10} p_{17} &
Y_{35}= p_1 p_4 p_7 p_{11} p_{13} p_{17} &
Z_{36}= p_9 p_{12} p_{13} p_{14} p_{17}
\\
X_{45}=p_1 p_3 p_4 p_6 p_8 p_{13} &
Y_{46}=p_4 p_8 p_{11} p_{12} p_{13} p_{15} &
Z_{41}=p_2 p_6 p_{12} p_{13} p_{14}
\\
X_{56}=p_3 p_8 p_9 p_{10} p_{12} p_{15} &
Y_{51}=p_2 p_7 p_{10} p_{11} p_{12} p_{15} &
Z_{52}=p_2 p_5 p_9 p_{12} p_{14}
\\
X_{61}=p_1 p_2 p_3 p_6 p_7 p_{10} &
Y_{62}=p_1 p_2 p_4 p_5 p_7 p_{11} &
Z_{63}=p_2 p_5 p_6 p_{14} p_{16}
\end{array}
\label{param}
\eeqa
The complete matrix of charges is
{\small
\beqa
{\tilde Q}=\pmatrix{
1 & 0 & -1& 0 & 0 & 0 & 0 & 0 & 0 & 0 & -1& 0 & 0 & 0 & 1 & 0 & 0 &|& 0\cr
0 & 1 & 0 & 0 & 0 & 0 & -1& 0 & 0 & 0 & 0 & 0 & 0 & -1& 0 & 0 & 1 &|& 0\cr
0 & 0 & 1 & -1& 0 & 0 & 0 & 0 & 0 & -1& 1 & 0 & 0 & 0 & 0 & 0 & 0 &|& 0\cr
0 & 0 & 0 & 1 & -1& 0 & 0 & 0 & 0 & 0 & 0 & 0 & -1& 1 & 0 & 0 & 0 &|& 0\cr
0 & 0 & 0 & 0 & 1 & 0 & 0 & 0 & -1& 0 & -1& 0 & 0 & 0 & 1 & -1& 1 &|& 0\cr
0 & 0 & 0 & 0 & 0 & 1 & 0 & 0 & 0 & -1& 0 & 0 & -1& 0 & 1 & -1& 1 &|& 0\cr
0 & 0 & 0 & 0 & 0 & 0 & 1 & 0 & 0 & -1& -1& 0 & 0 & 0 & 1 & 0 & 0 &|& 0\cr
0 & 0 & 0 & 0 & 0 & 0 & 0 & 1 & -1& 0 & 0 & 1 & -1& 0 & -1& 0 & 1 &|& 0\cr
0 & 0 & 0 & 0 & 0 & 0 & 0 & 0 & 0 & 0 & 0 & 1 & 0 & -1& -1& 1 & 0 &|& 0\cr
0 & -1& 0 & 0 & 1 & 0 & 0 & 0 & 0 & 0 & 0 & 0 & 0 & 0 & 0 & 0 & 0 &|&
\zeta_1\cr
-1& -1& 0 & 0 & 0 & 1 & 1 & 0 & 0 & 0 & 0 & 0 & 0 & 0 & 0 & 0 & 0 &|&
\zeta_2\cr
1 & 0 & -1& 0 & -1& 0 & 0 & 0 & 1 & 0 & 0 & 0 & 0 & 0 & 0 & 0 & 0 &|&
\zeta_3\cr
0 & 1 & 0 & 1 & -1& 0 & -1& 0 & 0 & 0 & 0 & 0 & 0 & 0 & 0 & 0 & 0 &|&
\zeta_4\cr
-1& 1 & 1 & 0 & 0 & -1& 0 & 0 & 0 & 0 & 0 & 0 & 0 & 0 & 0 & 0 & 0 &|&
\zeta_5\cr
} \nonumber
\label{thematrix}
\eeqa
}
where the first nine rows define $\IC^*$ actions that remove the
redundancy in the parametrization (\ref{param}), and the remaining five
implement the five independent $U(1)$ symmetries of the initial field
theory.

\begin{figure}
\centering
\epsfxsize=2in
\hspace*{0in}\vspace*{.2in}
\epsffile{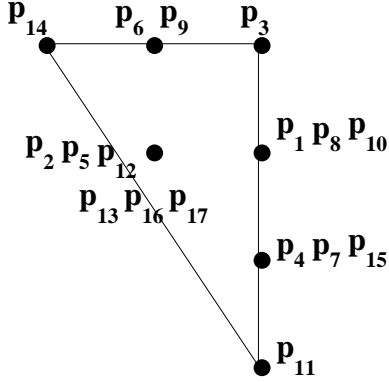}
\caption{\small The toric data for the moduli space of the field theory
for D3-branes at the $\IZ_2\times \IZ_3$ singularity.
}
\label{zsixp}
\end{figure}

When the blow-up parameters $\zeta_i$ vanish, the kernel of the matrix
above is
given by
{\small
\beqa
{\tilde T}=\pmatrix{
1 & 0 & 1 & 1 & 0 & 0 & 1 & 1 & 0 & 1 & 1 & 0 & 0 & -1& 1 & 0 & 0 \cr
0 & 0 & 1 & -1& 0 & 1 & -1& 0 & 1 & 0 & -2& 0 & 0 & 1 & -1& 0 & 0 \cr
1 & 1 & 1 & 1 & 1 & 1 & 1 & 1 & 1 & 1 & 1 & 1 & 1 & 1 & 1 & 1 & 1 \cr
}
\eeqa
}
The columns of this matrix provide the toric data describing the variety.
The polygon defined by the vector endpoints is shown in figure
\ref{zsixp}. As expected, it corresponds to the toric diagram of
\twothree.

This space can be described as a hypersurface in $\IC^4$. In order to see
that, define the invariant variables
\beqa
\begin{array}{ll}
x = p_1^4 p_2 p_3^6 p_4^2 p_5 p_6^3 p_7^2 p_8^4 p_9^3 p_{10}^4 p_{12}
p_{13} p_{15}^2 p_{16} p_{17}
& (=X_{12} X_{23} X_{34} X_{45} X_{56} X_{61}) \\
y = p_1 p_2 p_4^2 p_5 p_7^2 p_8 p_{10} p_{11}^3 p_{12} p_{13} p_{15}^2
p_{16} p_{17} & (=Y_{24} Y_{46} Y_{62})\\
z = p_2 p_5 p_6 p_9 p_{12} p_{13} p_{14}^2 p_{16} p_{17} &
(=Z_{14} Z_{41}) \\
w = p_1 p_2 p_3 p_4 p_5 p_6 p_7 p_8 p_9 p_{10} p_{11} p_{12} p_{13}
p_{14} p_{15} p_{16} p_{17} & (=X_{12}Y_{24}Z_{41})
\label{theinv}
\end{array}
\eeqa
which satisfy
\beqa
x y^2 z^3=w^6
\eeqa
This space is precisely \twothree, as shown in the main text, after
equation (\ref{def}).

\medskip

{\bf Some interesting resolutions}

Some interesting blow-ups of this singularity are mentioned in the main
text, and we provide here some details of the computations involved. Using
equation (\ref{param}), we see the blow-up associated to a vev for the
fields $Z_{63}$, $Z_{41}$ is obtained when the fields $p_2$, $p_5$, $p_6$,
$p_{12}$, $p_{13}$, $p_{14}$, $p_{16}$ are non-zero. This is achieved in
the region of FI space defined by$\zeta_1=-\zeta_4\ll 0$, $\zeta_2=\zeta_5=0$,
$\zeta_3\ll 0$, since the D-terms from the matrix (\ref{thematrix}) require
\beqa
|p_2|^2=|p_6|^2=|p_{14}|^2=-\zeta_3-\zeta_1, \quad
|p_5|^2=|p_{16}|^2=-\zeta_3, \quad |p_{12}|^2=|p_{13}|^2=-\zeta_1.
\eeqa
The basic invariants under the surviving $U(1)$ symmetries are
\beqa
x' & = & p_1^2 p_3^3 p_4 p_7 p_8^2 p_9 p_{10}^2 p_{15} \nonumber \\
y  & = & p_1 p_4^2 p_7^2 p_8 p_{10} p_{11}^3 p_{15}^2 p_{17} \nonumber \\
z & = & p_9 p_{17} \nonumber \\
w' & = & p_1 p_3 p_4 p_7 p_8 p_{10} p_{11} p_{15}
\label{moreinv}
\eeqa
Comparing with the original invariants (\ref{theinv}), the blow-up implies
the redefinition $x=x'^{\,2}z$, $w=w'z$. The remaining singularity is
therefore $x'y=zw'^3$.

An interesting further blow-up is achieved by vevs for the fields
$X_{45}$, $X_{23}$. Using (\ref{param}) this implies that we only keep
$p_7$, $p_9$, $p_{11}$, $p_{17}$ as dynamical, while all other variables
get non-zero vev. This vevs can be achieved in the region of FI space
defined by $\zeta_5=-\zeta_1-\zeta_4$, $-\zeta_3\gg \zeta_2\gg 0$,
$\zeta_4\gg -\zeta_1 \gg 0$. The D-term equations associated to the charge
matrix (\ref{thematrix}) are satisfied for
\beqa
\begin{array}{ll}
|p_1|^2=|p_4|^2=\zeta_1+\zeta_4 & |p_{10}|^2=|p_{15}|^2=\zeta_2 \\
|p_2|^2=|p_{14}|^2=-\zeta_1-\zeta_2-\zeta_3 & |p_{12}|^2 =-\zeta_1 \\
|p_3|^2=|p_8|^2=\zeta_1 + \zeta_2+\zeta_4 & |p_{13}|^2=\zeta_4 \\
|p_5|^2=-\zeta_2-\zeta_3 & |p_{16}|^2=-\zeta_3 \\
|p_6|^2=\zeta_4-\zeta_3 &
\end{array}
\eeqa

The variables (\ref{moreinv}) become
\beqa
x'=p_7p_9 \;\; ,\;\; y=p_7^2 p_{11}^3p_{17} \;\; ,\;\; z=p_9p_{17} \;\;
,\;\; w'= p_7 p_{11}
\eeqa
The new invariant is $y'=p_{11}p_{17}=y/w'^2$. The remaining singularity
is the conifold $x'y'=zw'$.

\section{Appendix: Tadpoles for the $\IZ_2\times \IZ_3$ orientifold}

As mentioned in the main text, the \twothree orbifold is actually
isomorphic to the $\IZ_6$ orbifold generated by the action on $\IC^3$
given by $z_i \to e^{2\pi i v_i} z_i$ with $v=(1,2,-3)/6$. The
non-compact orientifold studied in section 5.2 belongs to a large family
of models, introduced in \cite{abiu}, whose tadpoles can be computed in a
quite general fashion. So let us consider a general $\IZ_N$ orbifold, with
$N$ even, and generator defined by $v=(\ell_1,\ell_2,\ell_3)/N$, with
$\ell_1$, $\ell_3$ odd and $\ell_1+\ell_2+\ell_3=0$. We mod out this space
by the orientifold action $\Omega'=\Omega(-1)^{F_L}R_1R_2R_3$. We also
choose the Chan-Paton matrices $\gamma_{\theta,3}$ ($\gamma_{\theta,7_2}$)
to have $n_j$ ($m_j$) eigenvalues $e^{\pi i (2j-1)/N}$, and
$\gamma_{\theta,7_1}$ ($\gamma_{\theta,7_3}$) to have $w_j$ ($r_j$)
eigenvalues $e^{2\pi i j/N}$. The orientifold symmetry imposes
$n_j=n_{-j+1}$, $m_j=m_{-j+1}$, $w_j=w_{-j}$, $r_j=r_{-j}$.

The gauge group on the D3-branes is $\prod_{i=1}^{N/2}SU(n_i)$. The matter
multiplets can be obtained from the orbifold spectrum
\beqa
\sum_{\alpha=1}^3 \sum_{i=1}^N (\fund_i,\antifund_{i+\l_{\alpha}}) +
\sum_{i=1}^N \, [\, (\fund_i,{\ov m}_{i+\frac{\ell_1+\ell_3}{2}}) +
(\fund_i,{\ov w}_{i+\frac{\ell_2+\ell_3-1}{2}}) +
(\fund_i,{\ov r}_{i+\frac{\ell_1+\ell_2-1}{2}})
\eeqa
after imposing the identifications $\fund_i\equiv\antifund_{-i+1}$,
$m_i\equiv {\ov m}_{-i+1}$, $w_i\equiv {\ov w}_{-i}$, $r_i\equiv {\ov
r}_{-i}$ in the representations. When $i+\ell_\alpha=-i+1$ the bifundamental
$(\fund_i,\antifund_{i+\ell_\alpha})$ collapses to the two-index
antisymmetric (resp. symmetric) tensor representation $\Yasymm_{\,i}$
($\Ysymm_{\,i}$) when $\gamma_{\Omega',3}$ is symmetric (antisymmetric).

Let us turn to the tadpole computation, which can be obtained from the
appendix in \cite{afiv}. To simplify the notation, we define $s_{\alpha}
=\sin\pi kv_\alpha$, $c_\alpha=\cos\pi kv_\alpha$, ${\tilde s}_\alpha =
\sin 2\pi k v_\alpha$. Here we will be more sketchy than in appendix A,
for instance the volume dependences are not explicit, though they can be
extracted from the appropriate $\sin \pi k v_\alpha$ factors.

The cylinder tadpoles are
\beqa
\begin{array}{lll}
7_\alpha 7_\alpha\, : & \sum_k \frac{\textstyle 8 s_1s_2s_3}{\textstyle
4s_\beta^2 4s_\gamma^2} \; (\Tr \gamma_{\theta^k,7_\alpha})^2 &
\alpha\neq\beta\neq\gamma\neq\alpha \\
7_\alpha 7_\beta\, : & \sum_k 2\times \frac{\textstyle 2s_\gamma}{\textstyle
4s_\gamma^2}\; (\Tr \gamma_{\theta^k,7_\alpha}) (\Tr \gamma_{\theta^k,
7_\beta}) & \alpha\neq\beta\neq\gamma\neq\alpha \\
3\,\, 3\, : & \sum_k 8s_1s_2s_3 \; (\Tr \gamma_{\theta^k,3})^2 & \\
3\,\, 7_\alpha\, : & \sum_k 2\times 2s_\alpha \; (\Tr \gamma_{\theta^k,3})
(\Tr \gamma_{\theta^k,7_\alpha})
\end{array}
\eeqa
which can be neatly recast as
\beqa
{\cal C} = \sum_{k=1}^{N-1} \frac{1}{8s_1s_2s_3} [8s_1s_2s_3 \Tr
\gamma_{\theta^k,3} + \sum_{\alpha=1}^3 2s_\alpha \Tr
\gamma_{\theta^k,7_\alpha}]^2
\label{cylinders}
\eeqa

The Klein bottle tadpoles are
\beqa
16\sum_k \left[ \frac{8{\tilde s}_1{\tilde s}_2{\tilde s}_3}{4c_1^2 \,
4c_2^2\, 4c_3^2} - \frac{2{\tilde s}_2}{4c_2^2} \right]
\label{klein}
\eeqa
which can be written as
\beqa
{\cal K} = \sum_{k=1}^{N/2} \frac{1}{8{\tilde s}_1{\tilde s}_2{\tilde
s}_3} \; [\,32(s_1s_2s_3+c_1s_2c_3)\,]
\eeqa

Finally, the M\"obius strip tadpoles are
\beqa
\begin{array}{ccc}
3\, : & 8\sum_k 8s_1s_2s_3 \Tr (\gamma_{\theta^k\Omega',3}^{-1}
\gamma_{\theta^k\Omega',3}^T) & \\
7_\alpha\, : & 8\sum_k \frac{\textstyle 8c_\beta s_\alpha c_\gamma}{\textstyle
4c_\beta^2 4c_\gamma^2} \Tr (\gamma_{\theta^k\Omega',7_\alpha}^{-1}
\gamma_{\theta^k\Omega', 7_\alpha}^T) & \alpha\neq\beta\neq\gamma\neq\alpha
\end{array}
\eeqa
The orientifold requires the Chan-Paton matrices to satisfy 
\footnote{This follows from the results in \cite{bl} for D9- and 
D5$_i$-branes by a T-dualtiy along the three `internal' complex planes.}
\beqa
\Tr (\gamma_{\theta^k\Omega',3}^{-1} \gamma_{\theta^k\Omega',3}^T)
= \pm \Tr \gamma_{\theta^{2k},3} \quad ;\quad
\Tr (\gamma_{\theta^k\Omega',7_i}^{-1} \gamma_{\theta^k\Omega',7_i}^T)
= \mp\Tr \gamma_{\theta^{2k},7_i}
\eeqa
with the upper (lower) sign for the $SO$ ($Sp$) projection on the 
D3-branes. We also have
\beqa
\Tr \gamma_{\theta^N,3}=-1 \;\; ,\;\; \Tr\gamma_{\theta^N,7_2}=-1 \;\; 
,\;\; \Tr \gamma_{\theta^N,7_1}=1 \;\; , \;\; \Tr\gamma_{\theta^N,7_3}=1 
\eeqa
Using these properties, and after some algebra, the tadpoles can be recast as
\beqa
{\cal M} =  \mp\, 2 \sum_{k=1}^{N/2} \frac{1}{8{\tilde s}_1{\tilde
s}_2{\tilde s}_3}\; [32(s_1s_2s_3+c_1s_2c_3)]\; (8{\tilde s}_1{\tilde
s}_2{\tilde s}_3  \Tr \gamma_{\theta^{2k},3} + \sum_{\alpha=1}^3 2{\tilde
s}_\alpha \Tr \gamma_{\theta^{2k},7_\alpha})
\label{mobius}
\eeqa
with the upper (lower) sign for antisymmetric (symmetric)
$\gamma_{\Omega',3}$.

The tadpole cancellation conditions arising from (\ref{cylinders}),
(\ref{klein}), (\ref{mobius}) read
\beqa
& \prod_{\alpha=1}^3 2\sin \pi k v_\alpha \Tr \gamma_{\theta^k,3} +
 \sum_{\alpha=1}^3 2\sin \pi k v_\alpha \Tr \gamma_{\theta^k,7_\alpha}
+ &\nonumber \\
& \mp 32\delta_{k,0\,{\rm mod}\, 2} (\sin\pi k v_1\sin\pi kv_2 \sin\pi
kv_3 + \cos\pi k v_1\sin \pi kv_2 \cos\pi kv_3) & = 0
\eeqa
for all $k\neq 0$. It is a simple exercise to express the Chan-Paton
traces in terms of the integers $n_i$, $m_i$, $w_i$, $r_i$ and show that
these conditions are exactly equivalent to the cancellation of gauge
anomalies in the four-dimensional field theory on the D3-branes, described
above.

For the particular case of $v=(1,2,-3)/6$, the constraints read
\beqa
\begin{array}{l}
\Tr \gamma_{\theta,7_1} + \sqrt{3} + \Tr \gamma_{\theta,7_2}
-2 \Tr \gamma_{\theta,7_3} -2\sqrt{3} \Tr \gamma_{\theta,3} = 0  \\
\Tr \gamma_{\theta^2,7_1} + \Tr \gamma_{\theta^2,7_2} \pm 8 =0 \\
\Tr \gamma_{\theta^3,7_1} + \Tr \gamma_{\theta^3,7_3} =0 \\
\Tr \gamma_{\theta^4,7_1} - \Tr \gamma_{\theta^4,7_2} \pm 8 =0 \\
\Tr \gamma_{\theta^5,7_1} - \sqrt{3} + \Tr \gamma_{\theta^5,7_2}
-2 \Tr \gamma_{\theta^5,7_3} + 2\sqrt{3} \Tr \gamma_{\theta^5,3} = 0
\end{array}
\eeqa
which are satisfied by the models in section 5.2.

\end{document}